\documentclass[a4paper,11pt]{article}
\pdfoutput=1 % if your are submitting a pdflatex (i.e. if you have
\usepackage{jinstpub} % for details on the use of the package, please
\usepackage{subfigure}
\usepackage{graphicx}
\usepackage{xcolor}
                     
\usepackage{xspace}
\newcommand{\degree}{\ensuremath{^\circ}\xspace}

\newcommand{\htp}{\texorpdfstring{\ensuremath{\mathrm{H}_2^+}\xspace}{H2+}}

\title{\boldmath IsoDAR@Yemilab: A Report on the Technology, Capabilities, and Deployment}

\author[a]{J.R. Alonso}
\emailAdd{jralonso@mit.edu}
\author[a]{D. Winklehner}
\emailAdd{winklehn@mit.edu}
\author[d]{J. Spitz}
\author[a]{J.M. Conrad}
\author[i]{S.H. Seo}
\author[i]{Y.D. Kim}
\author[b]{M. Shaevitz}
\author[c]{A. Bungau}
\author[c]{R. Barlow}
\author[e]{L. Calabretta}
\author[f]{A. Adelmann}
\author[d]{D. Mishins}
\author[h]{L. Bartoszek}
\author[a]{L.H. Waites}
\author[i]{K.M. Bang}
\author[i]{K.S. Park}
\author[j]{E.A. Voirin}

\affiliation[a]{Dept.~of Physics, Massachusetts Institute of Technology, Cambridge, MA 02139, USA}
\affiliation[b]{Columbia University, New York, NY, USA}
\affiliation[c]{University of Huddersfield, Huddersfield, UK}
\affiliation[d]{University of Michigan, Ann Arbor, MI, USA}
\affiliation[e]{INFN Laboratori Nazionali di Legnaro, Legnaro, IT}
\affiliation[f]{Paul Scherrer Institute, Villigen, Switzerland}
\affiliation[h]{Bartoszek Engineering, Aurora, IL, USA}
\affiliation[i]{Institute for Basic Science -- Center for Underground Physics, Daejon, KR}
\affiliation[j]{eVoirin Engineering Consulting, Batavia, IL, USA}

\abstract{IsoDAR@Yemilab is a novel isotope-decay-at-rest experiment that 
has preliminary approval to run at the Yemi underground laboratory (Yemilab) in
Jeongseon-gun, South Korea.
In this technical report, we describe in detail the considerations for 
installing this compact particle accelerator and neutrino target system at the 
Yemilab underground facility. Specifically, we describe the caverns being prepared 
for IsoDAR, and address installation, shielding, and utilities requirements.
To give context and for completeness, we also briefly describe the physics 
opportunities of the IsoDAR neutrino source when paired with the Liquid Scintillator
Counter (LSC) at Yemilab, and review the technical design of the neutrino source.}

\begin{document}
\maketitle
\flushbottom

\section{Introduction}
\label{section:intro}

The IsoDAR (Isotope Decay At Rest) source offers a pure $\bar \nu_e$ flux 
from decay of $^8$Li that, when paired with the Liquid Scintillator Counter 
(LSC) detector at Yemilab~\cite{LSC}, will make precise and definitive measurements of the oscillations due to sterile neutrinos \cite{PRL} and test for many possible interactions 
beyond the Standard Model~\cite{elastic}, via the  inverse $\beta$-decay (IBD) $\bar \nu_e
+ p \rightarrow e^+ +  n$, and  {$\bar \nu_e$--$e^-$} elastic scattering (ES)
processes.

The source makes use of a cyclotron-accelerated beam, delivered to a novel decay-at-rest target-system, producing an isotropic flux of $\bar \nu_e$ from the $\beta$-decay of 839-millisecond $^8$Li. The system is described in detail below.
%Specifically, a high-intensity H$_2^+$ ion source feeds  a 
%60 MeV/amu cyclotron.   After acceleration,  the extracted H$_2^+$
%ions are stripped to form a proton beam.
%The proton beam is then transported to a
%target of $^9$Be, cooled by D$_2$O, producing neutrons.  The neutrons enter 
%a surrounding $\ge$99.99\% isotopically pure $^7$Li sleeve, where neutron
%capture results in $^8$Li, that $\beta$ decay with a half-life of 839 milliseconds.  
%The resulting high-intensity decay-at-rest (DAR) $\bar \nu_e$ flux, peaking at $\sim 6$ MeV, interacts in the multi-kiloton liquid scintillator detector at Yemilab, the LSC, allowing physics searches 
%exploiting the inverse beta decay (IBD) $\bar \nu_e
%+ p \rightarrow e^+ + n$, and  $\bar \nu_e$-$e^-$ elastic scattering (ES)
%processes.

\subsection{The Structure of this Report}
This report is one of several articles about the IsoDAR experiment 
and its implementation at the Yemilab underground facility. We first review the physics opportunities in section~\ref{sec:physics}, then discuss the required technology in section~\ref{sec:requirements} and then the technological advances in the accelerator and the target, in section~\ref{sec:cyclo}.

The main part of this technical report is a condensed version of 
our recent whitepaper: ``IsoDAR@Yemilab: A Conceptual Design Report for 
the Deployment of the Isotope Decay-At-Rest Experiment in Korea's New 
Underground Laboratory, Yemilab'' \cite{CDR} referred to below as the CDR.
As such, it provides a technical review of the
requirements and plans for installation at the Yemilab site, making these
aspects available for an interested general audience.
Further discussion on transport of equipment to the laboratory,
handling of equipment underground,  consideration of environmental and safety issues and
specifications of utilities can be found in the CDR.

\subsection{Physics Drivers for IsoDAR@Yemilab}
\label{sec:physics}
\begin{figure}[t!]       
\begin{center}
{\includegraphics[width=0.6\textwidth]{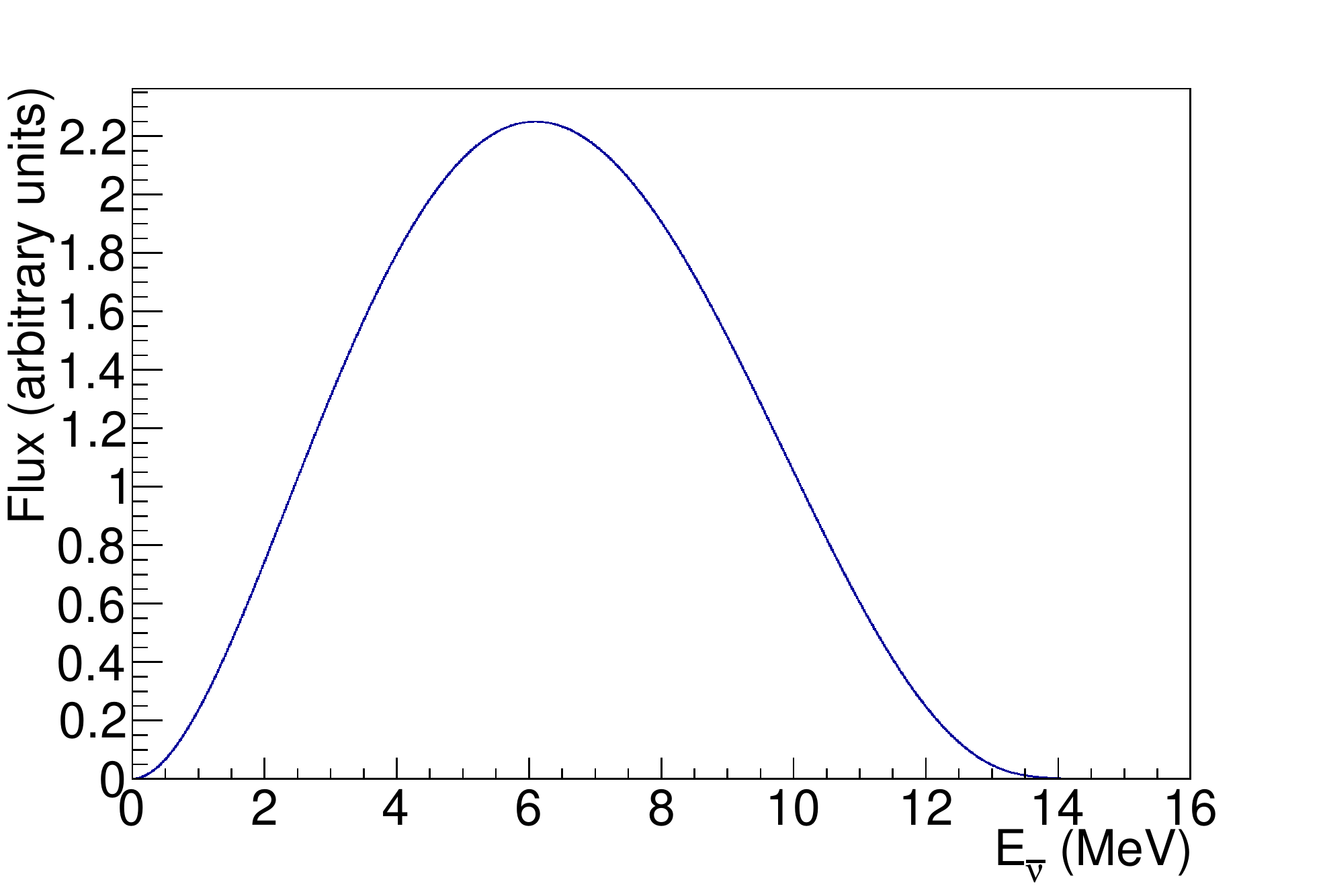}}
\end{center}  
\vspace{-1cm}
\caption{\label{fluxes} \footnotesize From \cite{Isophysics}. The antineutrino energy spectrum from the IsoDAR source.   
        }
\end{figure}

%% Interplay between Physics, Instrumentation and Installation
\begin{table}[b]
\centering
      \begin{tabular}{|c|c|} \hline 
Runtime  &  5 calendar years  \\ \hline
IsoDAR duty factor  &  80\%  \\ \hline
Protons on target/year  &  $1.97\times 10^{24}$  \\ \hline
$^8$Li/proton ($\overline{\nu}_e$/proton) &  0.0146  \\ \hline
$\overline{\nu}_e$/ 4 years  &  $1.15\times 10^{23}$  \\ \hline
1$\sigma$ uncertainty in $\overline{\nu}_e$ creation point  &  0.41~m  \\ \hline
IsoDAR@Yemilab mid-baseline  &   17~m  \\ \hline
IsoDAR@Yemilab baseline range  &   9.5-25.6~m  \\ \hline 
IsoDAR@Yemilab LS mass  &  2.26 ktons  \\ \hline
IsoDAR@Yemilab LS size (rad, height)  &  7.5 m, 15.0 m \\ \hline \hline
$\bar \nu_e + p\rightarrow e^+ + n $ (IBD) & $1.7\times 10^{6}$ events afer cuts\\ \hline 
$\bar \nu_e + e^- \rightarrow \bar \nu_e + e^- $(ES) & 7000 events after cuts\\ \hline
\end{tabular}
\caption{Assumptions for the physics case.  ``Cuts'' include fiducial volume and cosmic veto timing selection criteria. For ES, this also includes a visible energy requirement $E_{vis}>3$ MeV. See Ref.~\cite{Isophysics} for more information.}
\label{assumptions_table}
\end{table}
Our companion article ``Neutrino Physics Opportunities with the IsoDAR Source at Yemilab'' \cite{Isophysics} gives a detailed description of the neutrino physics program.
The physics capabilities are based on the parameters in Table~\ref{assumptions_table}.   
The primary IsoDAR physics program is made possible by the unique, single-isotope-produced ($^8$Li$\rightarrow ^8$Be$ + e^- + \overline{\nu}_{e}$), the relatively high energy ($<E_{\overline{\nu}_{e}}>$ $\sim 6$ MeV), and the pure $\bar \nu_e$ flux shown in Fig.~\ref{fluxes}. The immense (but isotropic) flux of antineutrinos ($1.15\times10^{23}$ in 4~years) that is produced in the target assembly, when placed as close as shielding will allow to the LSC, will provide very high statistics:  $1.7\times 10^6$ IBD events and 7000 ES events, after cuts. 

The neutrino physics measurements with IsoDAR@Yemilab, afforded by the 
expected high event rate, can be broken up into two categories: 
(1) a study of any oscillation-like behavior, possibly associated with one 
or more sterile neutrinos, which would present as a ``wave-like" IBD signal 
in terms of distance traveled ($L$), energy ($E$), and/or $L/E$ and 
(2) a search for non-standard neutrino interactions, which would present 
as a measured deviation from the well-predicted ES cross section. Both of
these physics opportunities are very well motivated, e.g., by the
3-5$\sigma$ anomalies indicating sterile neutrinos reported 
in numerous neutrino experiments~\cite{miniboone_new,MB_antinu,lsnd,reactor,source,2109.11482,IceCube}
and the simultaneous dearth of data in neutrino-based electroweak
scattering~\cite{global_reactor} and sensitivity to new physics in 
this sector. 

\begin{figure}[b!]       
\begin{center}
{\includegraphics[width=1.\textwidth]{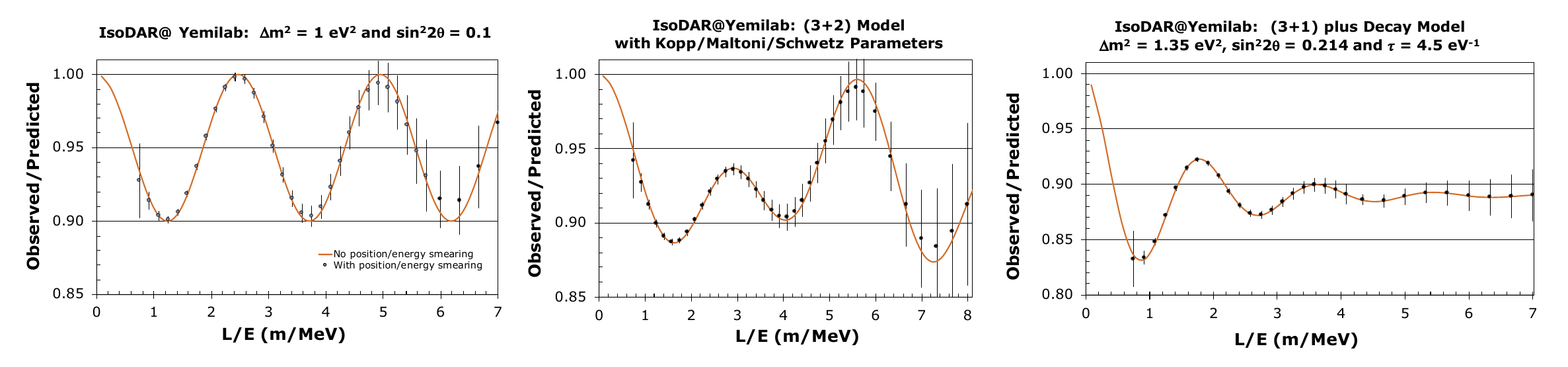}}
\end{center}  
\vspace{-1cm}
\caption{\label{wiggle} \footnotesize From \cite{Isophysics}. The IsoDAR@Yemilab ability to measure oscillations under three example new physics scenarios: a 3+1 model (left), a 3+2 model (middle), and a 3+1+decay model consistent with the 95\% allowed region observed at IceCube (right) \cite{MarjonThesis}.}
\end{figure}

\begin{figure}[t!]       
\begin{center}
{\includegraphics[width=0.6\textwidth]{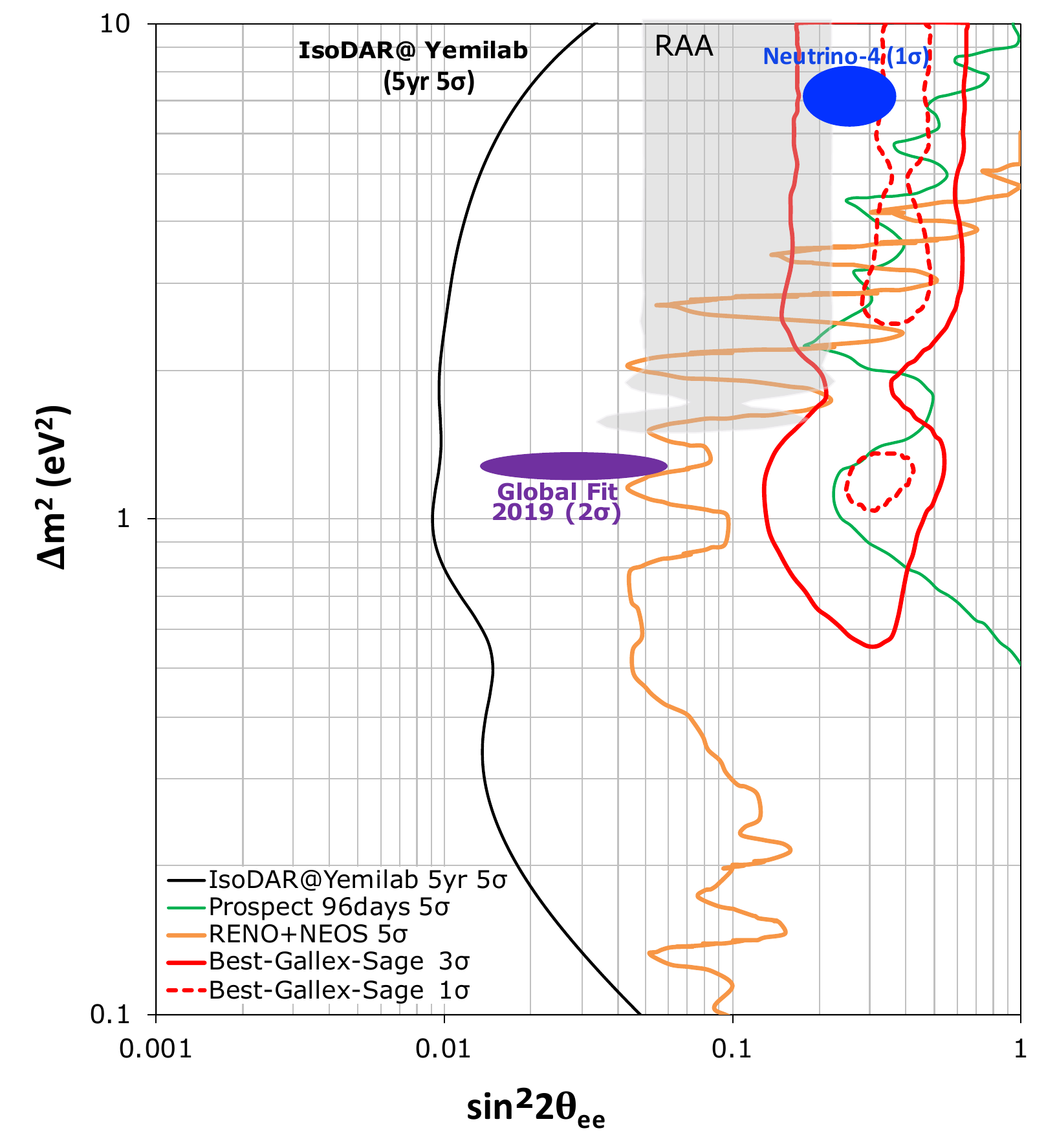}}
\end{center}  
\vspace{-1cm}
\caption{\label{osc_sens} \footnotesize  From \cite{Isophysics}. The 5$\sigma$ sensitivity of the IsoDAR@Yemilab experiment after 5 years of running.   
        }
\end{figure}

The IsoDAR@Yemilab  oscillation search will take advantage of the high rate, low background, and strong energy/vertex detector resolution for gaining unprecedented sensitivity to this new physics. The event rate as a function of $L/E$ for a few example oscillation signatures, including statistical uncertainties and smearing according to the expected energy and vertex resolutions, are shown in Fig.~\ref{wiggle}. The figure depicts oscillations in representative 3+1 (left), 3+2 (middle), and 3+1+decay (right) scenarios. The associated sensitivity within a 3+1 model is shown in Fig.~\ref{osc_sens}~\cite{Isophysics}. As can be seen, IsoDAR@Yemilab would comfortably provide world-leading sensitivity to oscillations involving a sterile neutrino. 

\begin{figure}[t!]       
\begin{center}
{\includegraphics[width=0.6\textwidth]{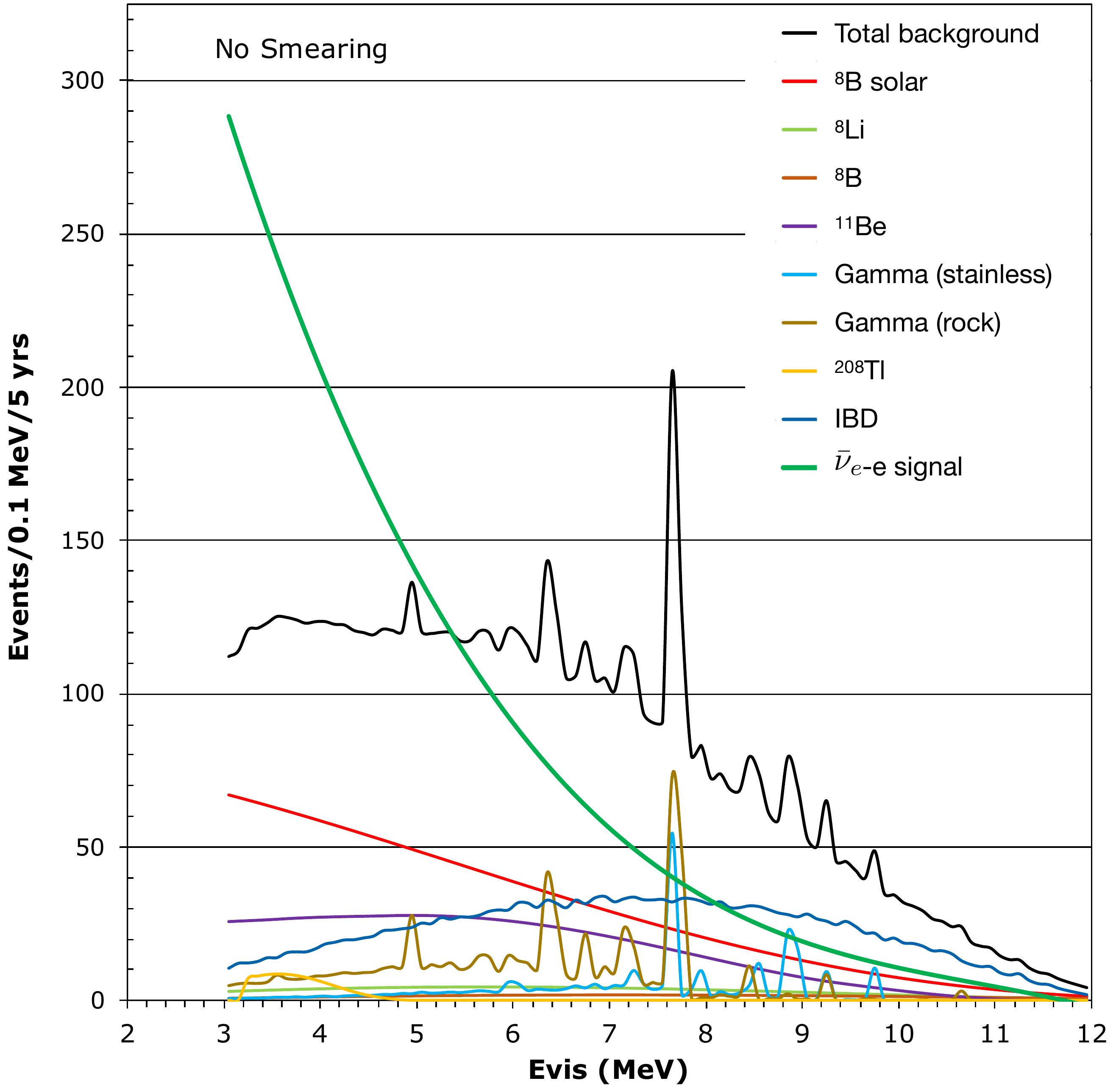}}
\end{center}  
\vspace{-1cm}
\caption{\label{es_rates} \footnotesize  {From \cite{Isophysics}. The ES signal/background event rates expected after 5~years of running IsoDAR@Yemilab.  
        }}
\end{figure}

Similarly to the IBD measurement, the IsoDAR@Yemilab ES study will take advantage of high rate, low background, and strong energy/vertex detector resolution in searching for this signature. Fig.~\ref{es_rates} shows the expected ES signal rate and backgrounds, coming from solar neutrinos, radiogenics, and cosmogenics. In terms of a weak mixing angle measurement, which can be thought of alternatively as the ``absence of non-standard neutrino interactions", we expect a $\delta \sin^2\theta_W$ sensitivity  of 1.9\%, using  rate  and  energy-shape  information  and  including  statistical  and  systematic  uncertainties, after 5 years of running. With the possible addition of directional reconstruction capabilities in the detector for mitigating signal/electron-like background, we can expect a sensitivity of 1.5\%~\cite{Isophysics}. This can be compared to the current reactor-based global fit of $\sin^2\theta_W=  0.252\pm0.030$, a 12\% measurement~\cite{global_reactor}. 

While the antineutrino flux provides the basis of the main physics IsoDAR@Yemilab program, interactions in the IsoDAR source also produce a high rate of neutrons and photons that are contained within the source region. These can be used for new physics searches involving conversion of photons and neutrons to non-standard particles.  
For example, Ref.~\cite{MHneutrons} provides examples of dark sector searches using a ``neutrons-shining-through-walls'' method. In addition, an interesting feature of the photon flux from the IsoDAR source that can be exploited towards gaining sensitivity to axions is the presence of mono-energetic lines from the decay of nuclear excited states near the target.

%\begin{figure}[t!]       
%\begin{center}
%{\includegraphics[width=0.6\textwidth]{weak_mixing_fixed.png}}
%\end{center}  
%\vspace{-1cm}
%\caption{\label{fluxes} \footnotesize  IsoDAR@Yemilab's sensitivity to $\sin^2{\theta_W}$ in comparison to past and future (DUNE~\cite{dune_weakmixingangle,dune_nd}) experiments, and a global reactor-neutrino analysis~\cite{global_reactor}. Aspects of this figure are adapted from Ref.~\cite{dune_weakmixingangle}.  
 %       }
%\end{figure}

\subsection{Requirements and Accelerator Technology Options}
\label{sec:requirements}
The selection of the accelerator type, and design decisions described below are the result of a careful study of requirements as well as capabilities and limitations of existing accelerator technologies.  This study can be found in reference~\cite{costeffective}; highlights are summarized here.

The fundamental requirements for the IsoDAR experiment are straightforward:  deliver a maximum rate of an optimal $\bar \nu_e$ spectrum into the fiducial volume of the LSC with minimum cost and maximum efficiency.  A cost-efficient package must be specified, and designed to fit in the underground environment adjacent to the large detector, placing emphasis on compactness, reliability, efficiency, and cost. Each of these points will be elaborated below.

\begin{itemize}

\item {\it $\bar \nu_e$ spectrum.}~~To provide a neutrino signal minimally affected by background, the neutrino energies should be above $\sim$4~MeV. Higher energy is also favored due to the increasing interaction cross section.  
Using $^8$Li as the parent isotope meets this requirement very well.  It provides a clean, well-understood, high-energy spectrum shown in Fig.~\ref{fluxes}, with an endpoint close to 14 MeV.

\item {\it Number of Events.}~~The effectiveness of a neutrino experiment is often ultimately related to the number of events that can be accumulated.  For IsoDAR, as the neutrino flux is emitted isotropically around the target, maximizing the number of events is determined by how close the target can be placed to the detector (maximizing the subtended solid angle of the fiducial volume), and by the number of neutrinos produced.  Closeness of the target is a careful balance between solid angle and shielding thickness to prevent neutrons produced in the target from penetrating to the fiducial volume.  This is discussed in Section~\ref{LSCshield}.  The optimum geometry provides about 5\% efficiency.  The neutrino flux produced from $^8$Li decays is ultimately related to the number of protons striking the neutron-producing target. Yield calculations have been based on 10~mA of 60~MeV protons striking the target.

\subitem {\it Proton energy.}~~Fig.~8 and Table 5 of the Cost Effectiveness
study~\cite{costeffective} address the neutron production for protons 
striking a target versus energy (result: a straight line on log-log plot;
empirically, [$(n/p) \approx 3\times10^{-5}\times\mathrm{E}^2$], 
where E is in MeV).  The obvious conclusion is that higher energies 
produce substantially more neutrons per proton. Specifically, at 60~MeV 
the (n/p) ratio is about 0.11, while at 1~GeV the ratio is about 30.  
But the footprint of a 1~GeV Spallation Source would not fit underground.
Fig.~10 of \cite{costeffective} addresses the maximum current of 
different accelerator technologies, and clearly shows that isochronous
cyclotrons are the most efficient.  They can also be compact, making them the clear best choice for an underground
installation. 

\subitem {\it Beam current.}~~Compact cyclotron designs are very mature, with demonstrated reliability; driven by the medical isotope industry.  Current limits for these isotope accelerators are about 1~mA to 2~mA. These limits are driven by beam loss in the central region (at injection energies of a few 10’s of keV), and by extraction stripper foil lifetimes due to heating from convoy electrons generated in the stripping of the H$^-$ ions accelerated~\cite{EJNMMI}.  As discussed in section~\ref{accel}, our 
technical innovations are expected to allow reaching the 10~mA beam current requirement in a compact cyclotron.

\item {\it Alternate accelerator technologies.}~~Other than isochronous cyclotrons, compact or separated sector, no other circular accelerator technology could come close to the 10 mA current specification.  A separated-sector isochronous-cyclotron based system could undoubtedly be designed using the same features as the compact-cyclotron system being adopted, however the system would be larger.  

%For example, the 
%PSI Injector 2 is such a separated sector machine~\cite{Injector2}, its outer diameter is 8 meters across, but it accelerates protons (to 72 MeV), not \htp.  \htp is important to increase the current from the present maximum of 3 mA to 10 mA, but the increased rigidity would make the outer diameter even bigger.  It would be difficult to fit into a cavern that would easily accommodate the compact cyclotron.

A linear accelerator could produce the required peak beam current, but would need to be superconducting to run at the 100\% duty factor to achieve the integrated beam power on target.  Even if superconducting, a structure to produce 60 MeV protons would be long, the low final energy requiring distributed quarter-wave resonators, so the {\it engineering} energy gain per meter is low.  Fitting such a linac in the Yemilab caverns would be difficult.  The cost of such a structure, and added complexity of a cryogenic plant makes such a system unattractive.  Ref.~\cite{costeffective} concludes the cost would be at least double the cost of a cyclotron system.

\end{itemize}

The conclusion of our study is that a compact isochronous cyclotron presents the optimal technology to be pursued for the IsoDAR experiment.  
In the following section we summarize the elements of the system, 
and progress towards the final design and implementation.

\section{Review of the Technical Design and R\&D Establishing the IsoDAR Source}
\label{sec:cyclo}

The elements of the IsoDAR $\bar \nu_e$ source are as follows:

\begin{figure}[t!]
\centering
\includegraphics[width=4.5in]{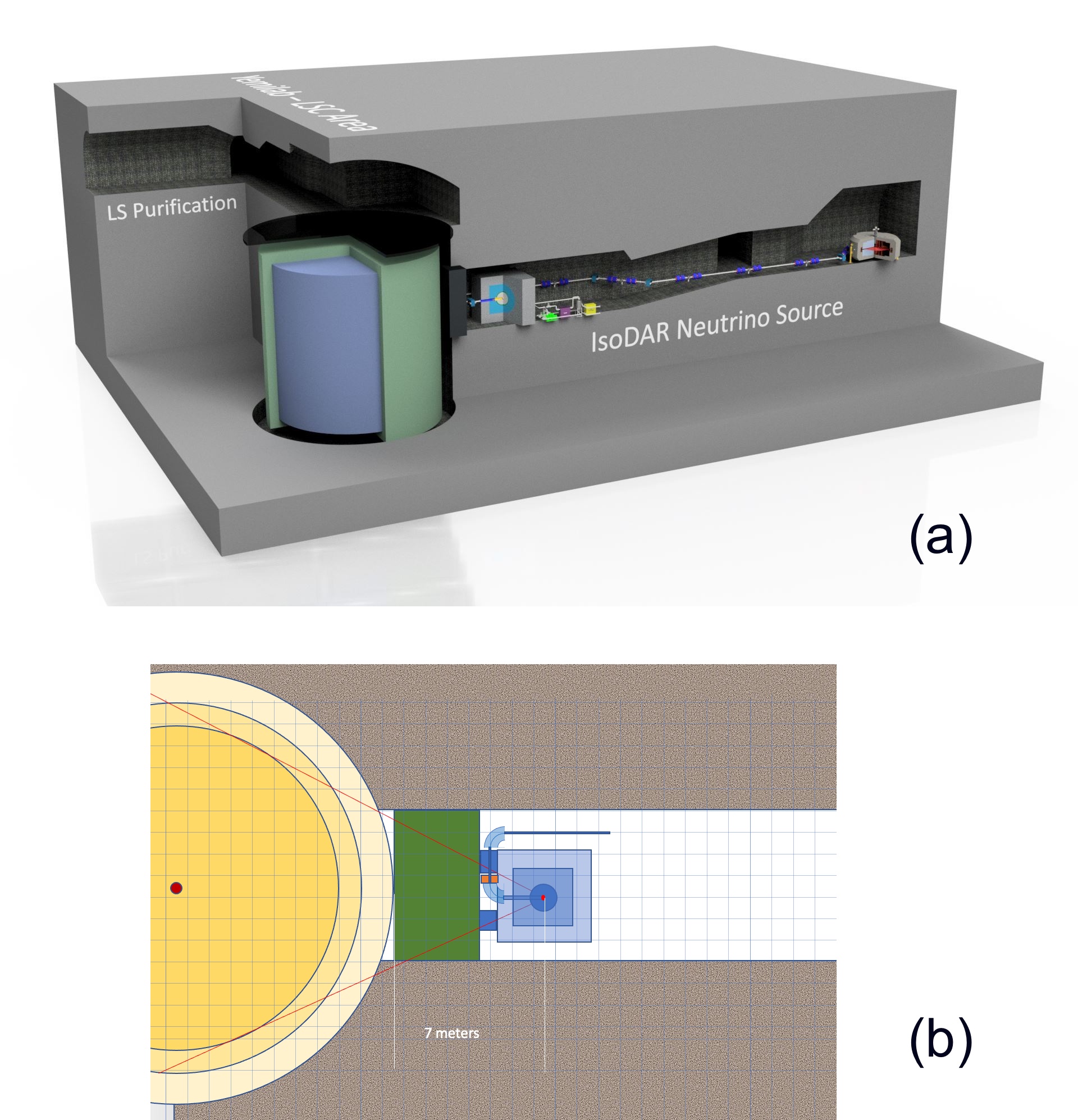}
\caption{
\footnotesize Schematic of the IsoDAR experiment deployed at the Yemilab site; (a) shows the cyclotron
at the far right corner and the transport line taking the beam through to the target area. The target assembly is represented by the blue cubes (steel and concrete) in (b), the target itself is the small red dot at the center.  The sleeve with the $^8$Be + $^7$Li where the $\bar \nu_e$ flux is produced, surrounds the target. The beam line comes in via two 90$\degree$ bends so the beam strikes the target pointing away from the detector. This greatly reduces the fast neutron flux directed towards the detector. The target volume of the detector is represented by the blue cylinder in (a) and darker yellow in (b), and the buffer and veto regions are shown in green (lighter yellow in (b)). }
\label{deployment}  
\vspace{0.2in}
\end{figure}

\begin{itemize}
\itemsep-0.1em
\item An ion source producing a $\sim$10~mA DC beam of \htp ions.
\item A Radio-Frequency Quadrupole (RFQ) buncher that allows $>60$\% capture of the \htp beam in the cyclotron.
\item An axially-injected compact cyclotron that produces an extracted beam of 5~mA of \htp ions at 60~MeV/amu.
\item A stripper foil close to the extraction point that converts the 
5~mA of \htp to 10~mA of protons, followed by a dipole magnet to direct 
the stripped protons into the transport line.
\item A transport line (MEBT) that brings the protons to the target area, then through two 90\degree bends onto the target.
\item A wobbler magnet that spreads the beam uniformly over the face of the target.
\item A layered beryllium and heavy-water target that is struck by the proton beam, producing large quantities of
neutrons.
\item A sleeve containing a mixture of $\ge$99.99\%-enriched $^7$Li ($\sim25\%$) and beryllium ($\sim75\%$) that 
is flooded by these neutrons, which are moderated and captured to make
the parent $^8$Li.  
The subsequent $\beta$-decay of the $^8$Li (839~ms half-life)
produces the electron-antineutrinos. Note, the beryllium in the sleeve helps multiply the neutron flux.
\end{itemize}

\begin{figure}[t!]
\centering
\includegraphics[width=2.5in]{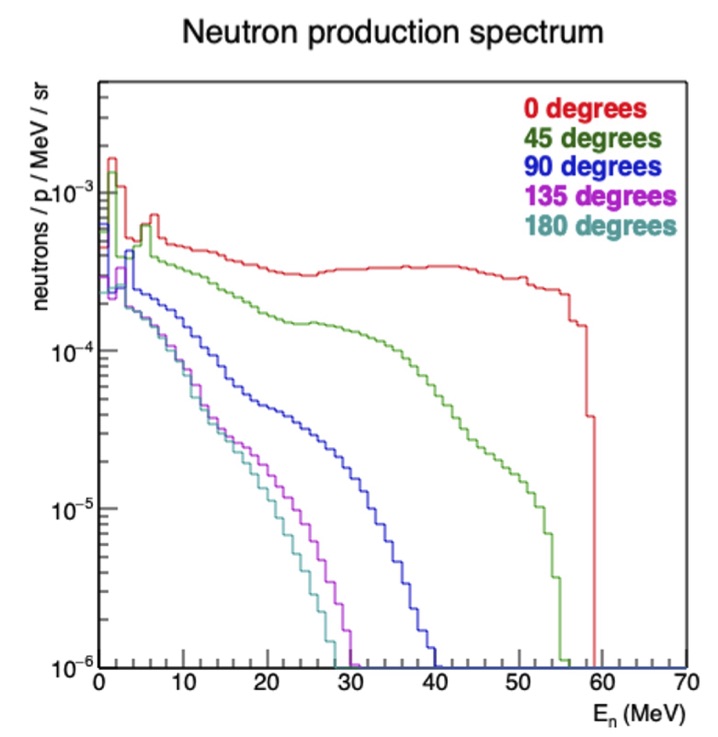}
\caption{{\footnotesize Angular distribution of neutrons emerging from the target.  High-energy neutrons in the 
backwards direction are greatly attenuated, reinforcing the value of the beam entering the target going away from
the detector.}
\label{neutron_spetrum}}
\vspace{0.2in}
\end{figure}

The layout of these components in the Yemilab setting is shown in Fig.~\ref{deployment} (a). 
The cyclotron is placed in the dedicated Cyclotron Room at a bend in the Entrance Ramp to the Target Room. (See also Fig.~\ref{IsoYemiLayOut} for a plan view of the layout.)
%The stripper and analysis magnet are close to the extraction point from the cyclotron.  
%The transport line or MEBT (Medium Energy Beam Transport) brings the beam down the ramp and to the target.
A large steel and concrete block (shown in green in 
Fig.~\ref{deployment} (b)) covers the opening to the LSC detector, 
to shield the detector from gammas and neutrons
generated in the Target Room.  The two 90$\degree$ bends in the beam line orient the beam so it strikes
the target going away from the detector.  Fig.~\ref{neutron_spetrum} demonstrates the value of this concept: the flux of
high-energy neutrons emitted in the backward direction, towards the detector, is greatly reduced
in both energy and intensity.
The target of nested shells of beryllium cooled with D$_2$O stops the beam and produces 
the neutrons shown in 
Fig.~\ref{neutron_spetrum}.  These neutrons are moderated as they stream into the sleeve surrounding the target where
they are captured by the $^7$Li.  The high fraction $~$(75\%) of beryllium powder in the sleeve has been optimized to multiply the neutrons, and thus maximize the yield of $^8$Li \cite{bungau:shielding}. The target and sleeve are surrounded with highly efficient shielding material to minimize neutrons emerging from the shielding
structure.

As the purpose of this section is to provide context for the layout underground and the infrastructure requirements, the elements are discussed 
only briefly.  For more substantial information on the components, please see the 2015 Technical Facility CDR on the neutrino source \cite{CDR2015}.  Updates to that information and substantial detail on the most complex elements of the neutrino source are provided in references ~\cite{winklehner:nima, winklehner:mist1, winklehner:RFQDIP, bungau:optimization, bungau:shielding}

\subsection{Overview of the Accelerator System}
\label{accel}

IsoDAR will deliver an order-of-magnitude more beam current than existing compact cyclotrons.
 The standard commercial compact cyclotrons today are limited by space-charge forces, especially at injection energies.
One must compress the charged particles into a tight bunch for injection and acceleration into the cyclotron, 
but the charge in this bunch generates a strong repulsive force (referred to as ``space charge'') 
which at high beam currents becomes too large to be overcome by the available focusing forces.  
A second limit for conventional cyclotrons,  all of which accelerate H$^-$, is the lifetime of the stripping foil
used for extracting the beam.
 
To reach our performance goal, the IsoDAR system must address both of these
challenges, and must pay great attention to efficiently bunching, injecting,
accelerating and extracting the very high-current beam.  
In all of this, it is imperative to minimize beam loss.
First of all, large beam losses require higher currents from the ion source 
to ensure adequate beam intensity on target. 
Secondly, beam losses for high intensity beams can damage the accelerator.  
At the low energies at the start of acceleration in the central region of 
the cyclotron, beam loss causes sputtering of material and voltage breakdown; 
at high energies, beam loss into the walls of the cyclotron produces neutrons
that cause activation and severely limit the ability to perform maintenance 
on the cyclotron because of the high radiation fields.

Three important design breakthroughs will allow us to reach the necessary intensities.   

\begin{figure}[tb!]
\begin{center}
\vspace{-0.2in}
{\includegraphics[width=5in]{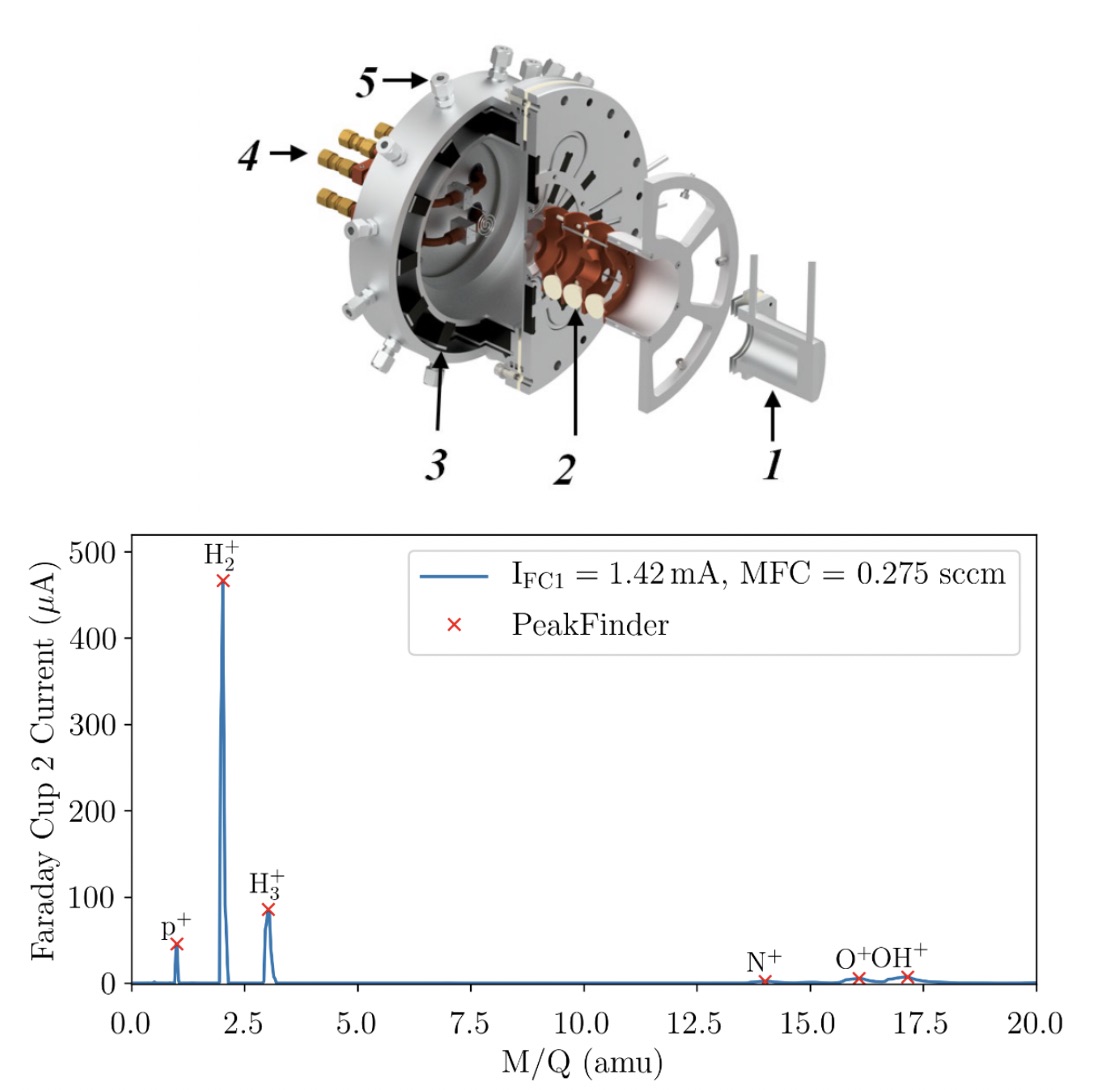}}
\vspace{-0.2in}
\end{center}  
\caption{\label{MIST-1}  Cutaway of MIST-1, filament-driven H$_2^+$ source, and recent spectrum.  H$_2^+$ peak dominates 
over proton and H$_3^+$, contaminant fraction is 
low \cite{winklehner:mist1}.}
\end{figure}

First, we use \htp ions rather than protons, which has benefits at a number
of places in the acceleration cycle, not the least being a reduction 
in space-charge effects.
A prototype high-current, filament-driven multi-cusp ion source has 
been built, shown in Fig.~\ref{MIST-1} that also shows the results of 
first studies of the source~\cite{winklehner:mist1}. 
Development and optimization of this source is continuing; at this 
writing \htp current density and 
ionic fraction are moving well towards design
requirements~\cite{winklehner:ICIS2021}.

\begin{figure}[t!]
\begin{center}
\vspace{-0.2in}
{\includegraphics[width=5in]{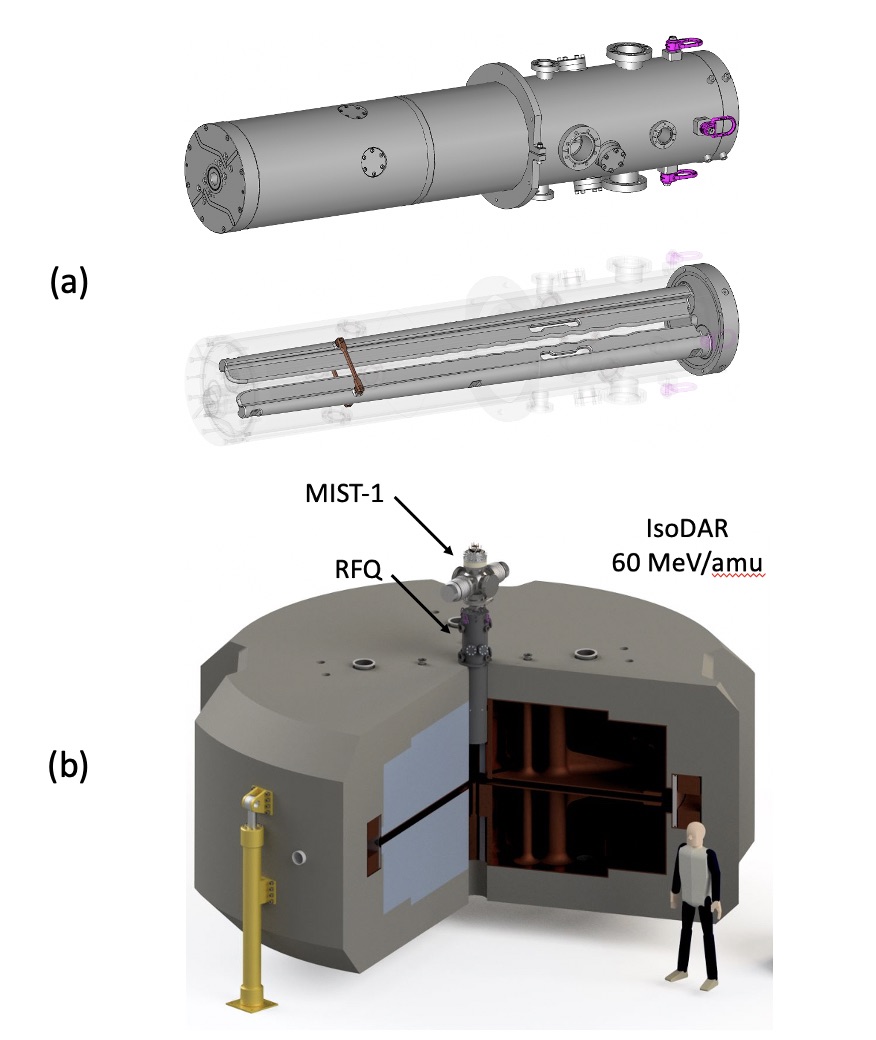}}
\vspace{-0.2in}
\end{center}  
\caption{\label{RFQ}  (a) Engineering model of RFQ designed by Bevatech GmBH, Frankfurt.  The bottom half of (a) shows two vanes of the split-coaxial structure mounted on one end plate.  The other two vanes are mounted on the opposite end plate. This geometry allows for the very low resonant frequency of 32.8 MHz, the frequency driving the cyclotron RF system.  The MIST-1 source is mounted on the right side, beam exits from the left side. (b) The RFQ is shown mounted along the central axis of the cyclotron.  A bit more than half of the structure is inside the steel of the cyclotron magnet.  The exit point is located 20 cm from the first accelerating Dee of the cyclotron.  This short flight path preserves most of the bunching provided by the RFQ.}
\end{figure}

Secondly, we employ direct axial injection through an RFQ
buncher for the first time in a compact cyclotron, allowing the high
efficiency beam capture that is necessary to reach high intensity~\cite{winklehner:RFQDIP}.
This RFQ replaces the conventional ``low energy beam transport'' (LEBT) that 
transports the continuous stream of particles from the ion source to the point 
of cyclotron injection through the so-called ``spiral inflector''~\cite{winklehner:spiral}.
For efficient capture, the cyclotron can accept only beam 
that is within  $\sim \pm10^\circ$ of the synchronous phase
of the RF accelerating voltage.
As beam from the ion source is continuous, capture efficiency is less than 10\%.
This requires a factor of 10 higher current from the ion source, and leads to large beam losses in the central region of the cyclotron.  
%As stated above, this beam loss leads to sputtering damage 
%and high-voltage arcing problems.
Because of the high space-charge in low-energy high-current beams, ``classical'' double-gap bunchers, 
attempting to compress more particles into the acceptance window,
 only increase the capture efficiency to about 30\%.
The 1-meter long RFQ buncher, operating at the same frequency as the cyclotron, is capable of placing
over 60\% of the continuous beam from the ion source into the phase-acceptance window of the cyclotron and so to increase beam intensity. 
The RFQ buncher is shown in Fig.~\ref{RFQ}. 
Construction of a prototype RFQ for IsoDAR is now underway \cite{winklehner:nima}. 

\begin{figure}[tb!]
\begin{center}
\vspace{-0.2in}
{\includegraphics[width=3.in]{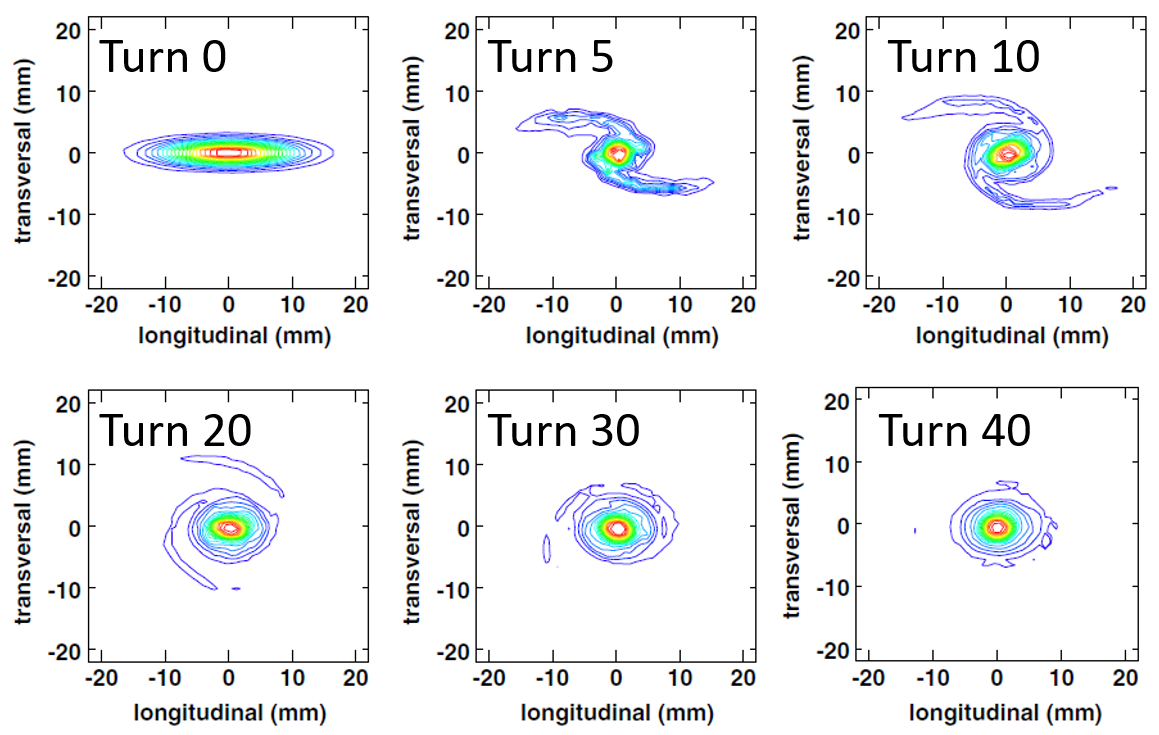}}
\vspace{-0.2in}
\end{center}  
\caption{\label{vortex}  Vortex motion in PSI Injector II; simulations using the OPAL code confirm the 
experimental observations \cite{yang:vortex}. In only a few turns the bunch evolves into a spiral, 
with the majority of the particles wrapped into a tight bunch.  Halo particles, amounting to 10-20\% of the total, 
are scraped off with judiciously placed collimators, all before the beam has reached 5 MeV, so causing no activation.  
From turn 40 onwards there is virtually no change in the bunch shape.}
\end{figure}

Thirdly, and perhaps the most important breakthrough, was the discovery 
that for high-current beams in cyclotrons, where space charge is a dominant force,
an effect called ``vortex motion'' can stabilize the beam and reduce growth 
during acceleration. This surprising finding was first observed at the 
PSI Injector II cyclotron \cite{stetson:vortex, stammbach:vortex}.  
As shown in  Fig.~\ref{vortex}, as the beam circulates,
the combination of space charge effects and external focusing forces 
induce bunches to curl tightly in phase-space.
This prevents the massive losses that some accelerator 
physicists had predicted would occur on the extraction septum,
due to radial overlap of bunches in adjacent turns close to extraction 
from the cyclotron \cite{yang:vortex, winklehner:60mev}.
The formation of the stable vortices does push a fraction of the bunch into larger ``halo'' orbits, but these can be removed using carefully-placed collimators, all located in the central region where the beam energy is below the Coulomb barrier so no activation or neutrons are produced.

\begin{figure}[tb!]
\begin{center}
\vspace{-0.2in}
{\includegraphics[width=0.95\textwidth]{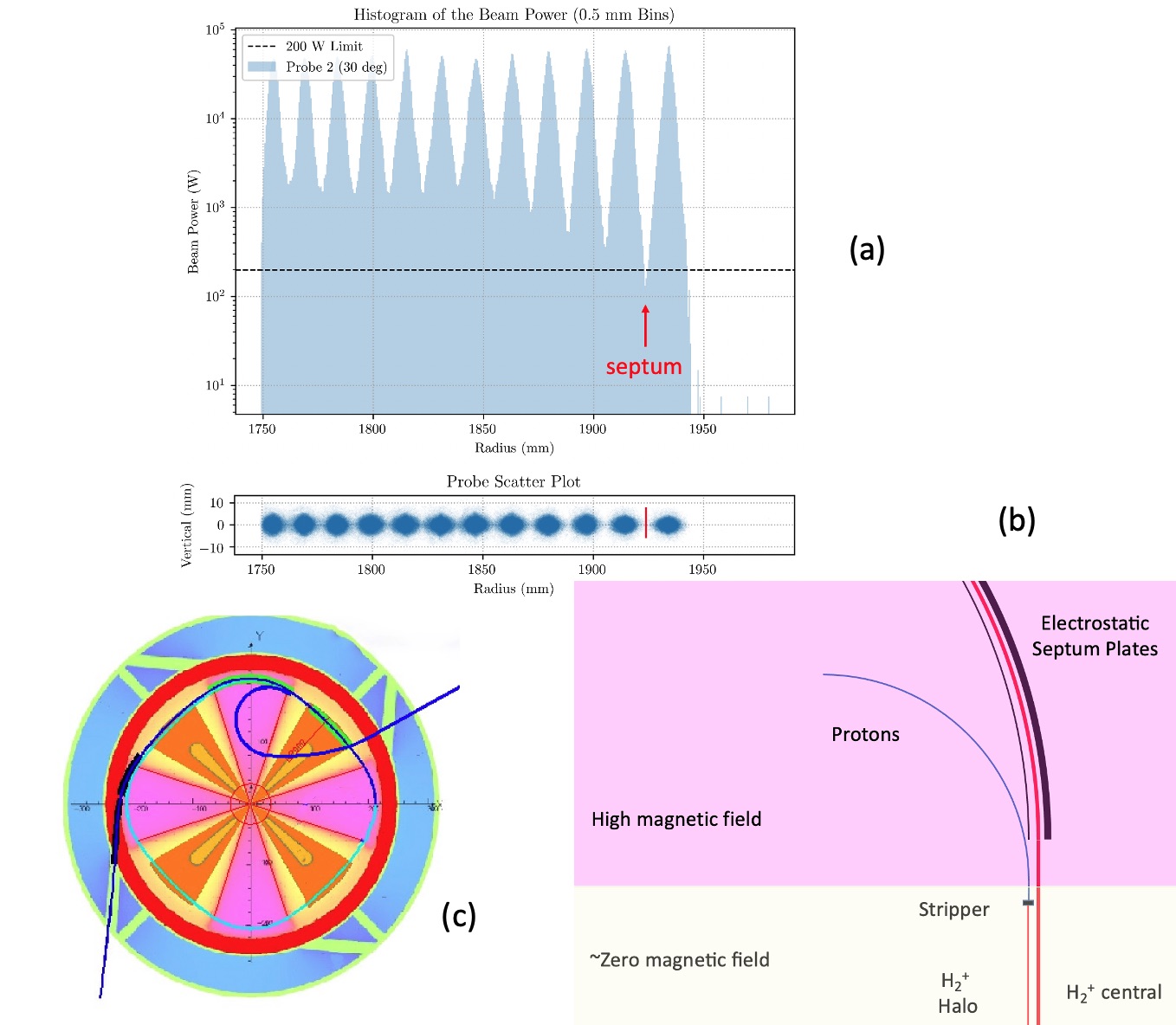}}
\vspace{-0.2in}
\end{center}  
\caption{\label{extract}  (a) Simulations of outer orbits in the cyclotron showing clean turn separation, particularly on the last turn.  Note the logarithmic scale.  (b) Schematic of the electrostatic extraction channel (black) and the H$_2^+$ beam being extracted (red).  Residual beam between turns, referred to as ``halo'' that would strike the inner septum plate is intercepted by the thin stripper.  These ions are converted to protons that are more sharply bent in the high ``hill'' magnetic field of the cyclotron, missing the septum plate.  (c) Plan schematic of cyclotron showing the four high-field ``hills'' (pink), the four low-field ``valleys'' (yellow), the orange RF accelerating Dees, the red coil surrounding the entire pole section, and the blue yoke return steel for the magnet.  The H$_2^+$ extraction channel is shown with two electrostatic deflectors followed by two magnetic channels.  The orbits of the stripped protons are also shown as the blue spiral exiting to the upper right.}
\end{figure}

%Another feature of using H$_2^+$ ions relates to extraction from the cyclotron, which is described below.
Instead of extracting the beam with a stripping foil, we revert to
the early technique of an extraction channel employing a thin electrostatic septum for guiding
the beam out of the cyclotron.  
For this to work, one must have good turn-to-turn separation at the last orbits, and must
keep the size of the bunch small in order to have few particles at the radius where the septum
is located.  Introducing a structure resonance helps increase turn separation at the extraction
point ~\cite{seidel:extraction}, while the above-mentioned vortex effect controls the bunch size.
The result, shown in Fig.~\ref{extract} shows that the loss on the septum will be less than 100 watts which,
considering the total beam power is 600 kilowatts, is a remarkable accomplishment.  This level
of beam loss is well below the ``rule-of-thumb'' from PSI, of maintaining beam losses below 200 watts
to enable hands-on maintenance of accelerator components.

%But there is a further step we can take to protect the electrostatic extraction septum.  
Placing a narrow stripping foil just upstream of the septum intercepts ions that would strike
the septum.  These H$_2^+$ ions are converted to protons which are then bent inwards by the 
cyclotron magnetic field, and miss the septum.  They orbit tightly in the strong ``hill'' 
section of the magnetic field, and emerge in the weaker ``valley'' on a trajectory that takes
them safely out of the cyclotron. The details of this extraction scheme are shown in Fig.~\ref{extract}.

% The hill/valley design of the magnet may be unfamiliar to some readers.  This is employed in cyclotrons that are ``isochronous,'' ({\it i.e.} the time for a particle to complete one orbit, or ``turn'' is constant, and does not depend on the energy of the particle). In such machines, relativistic effects are compensated by a radially increasing  turn-averaged field, and vertical focusing is achieved by the azimuthally varying field (AVF) generated by the hills and valleys. This allows for a single-frequency RF system  to be used, and for particles of all energies, from injection to extraction, to reside stably in the cyclotron.  Thus, particles are continuously accelerated and extracted.

%Details of the transport and assembly of the cyclotron components to the Cyclotron Room at Yemilab are discussed later in this paper.

%The accelerating system will be delivered in parts and assembled in the Yemilab space.  Studies in Ref.~\cite{CDR2015} consider limitations on size that are more restrictive than what is required at Yemilab. In particular, we are confident that 
%the cyclotron coil, which is 5 m in diameter, can be brought through all Yemilab passages when tilted on the diagonal.   Therefore, this complexity will not interfere with the goals of experiment but does add time to construction.

\subsection{Beam Transport (MEBT) System}
Figs.~\ref{deployment}~and~\ref{target_room} show the transport line connecting the cyclotron to the target referred
to as the MEBT (Medium Energy Beam Transport).

The H$_2^+$  ions extracted from the cyclotron are passed through a thin
carbon stripping foil that removes the electron, leaving two bare protons.
%It is easier and safer to transport protons: beam losses can be better controlled.
The stripping system and an analyzing magnet are placed close to the extraction point.  The analyzing magnet  
steers protons into the transport line, and serves to monitor the integrity of the stripping foil.  A detector placed at the orbit of unstripped H$_2^+$ ions picks up impending failure of the foil. A chain mechanism quickly brings a new foil into position.

Beyond the stripping stage, the proton beam is transported down the ramp to the target area.
Standard transport elements are used: quadrupole magnets for focusing and dipoles
for bending the protons. Beam position and profile are monitored by standard beam
diagnostic devices designed for high-current beams.

Collimation stages at necessary points along the line remove halo from the beam, these ``controlled'' beam loss points are enclosed in localized shielding to attenuate the neutrons produced from stopped particles.

The remaining ``uncontrolled'' loss is from scattering of protons in the residual gas in the beam line, which occurs along the length of the line.  The scattering cross section for protons is substantially less than for \htp ions, justifying the stripping of these ions as soon as possible after the extraction point.  This loss is mitigated by specifying the highest practical vacuum in the beam line.

As described earlier, the two 90\degree bends bring the beam onto the target going away from the LSC, decreasing the flux of fast neutrons pointing towards the LSC.  

The beam is spread out with a wobbling magnet to paint the beam over the face of the target, making the distribution of power over the target as even as possible.  

\subsection{Target System}{\label{target-section}}
\begin{figure}[b!]
\centering
\includegraphics[width=3.5in]{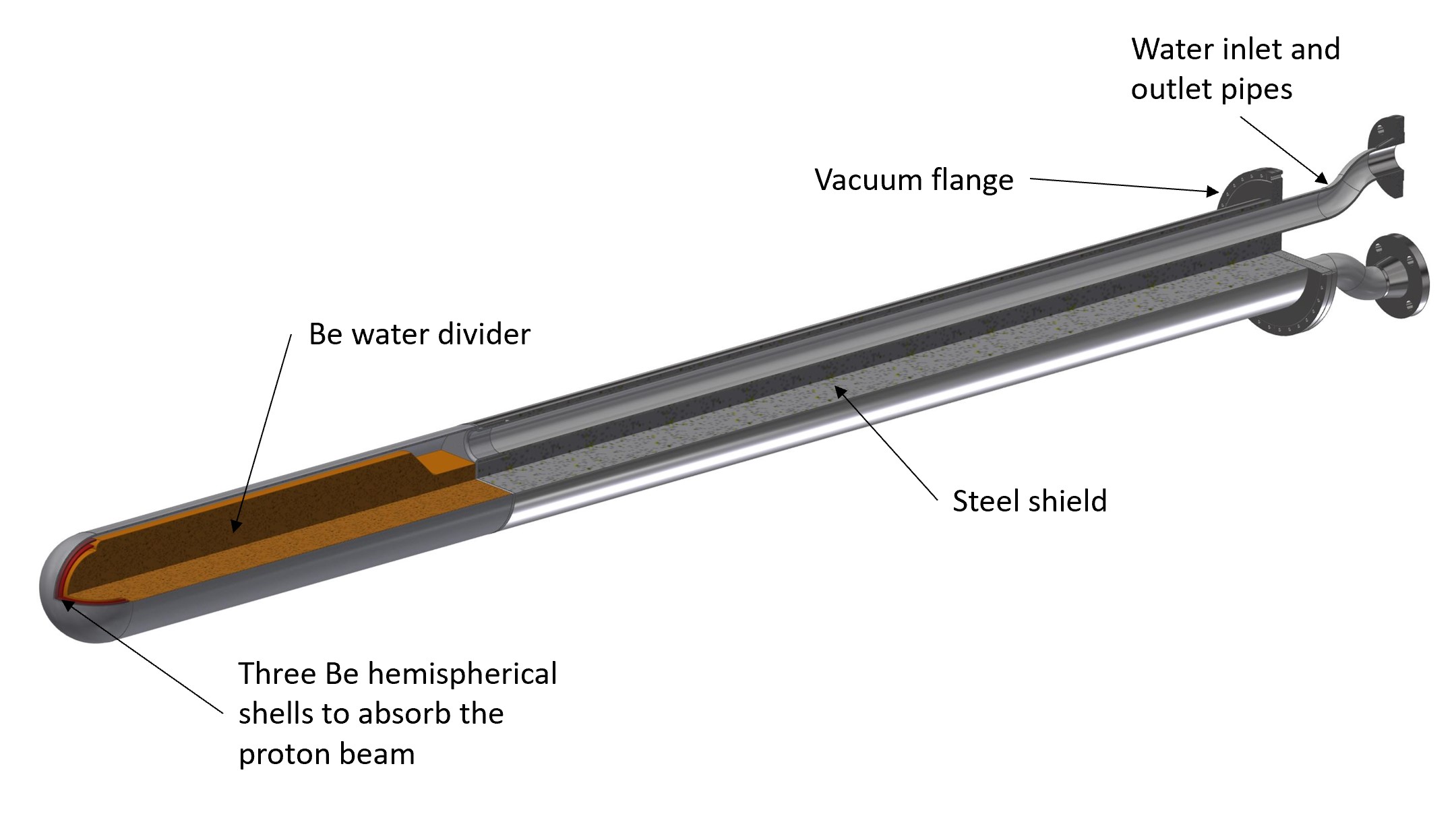}
\caption{{\footnotesize  Referred to as the ``Torpedo,'' the target assembly consists
of nested beryllium hemispheres separated by cooling water (D$_2$O) channels.  The entire assembly
can be easily detached from the cooling lines and removed into a shielded storage area.}
\label{torpedo}}
\vspace{0.2in}
\end{figure}
The target assembly, referred to as the ``torpedo'' (see Fig.~\ref{torpedo}), 
is designed for easy removal and replacement from the back of the target shielding
block.  Fig.~\ref{shielding} shows the shielding structure surrounding the target 
and sleeve, with a 20 cm (inner diameter) vacuum pipe running the full length.  The 
inlet side brings the beam onto the target face.  The back side allows the torpedo
to be slid out.  
As mentioned earlier, the orientation of the target assembly in the hall faces away
from the detector (See Fig.~\ref{deployment} (b)), 
and so provides ample room for removal of the long torpedo assembly.
This will probably be done remotely, using a robot, because of high radiation levels.
The spent targets will be stored in bore holes drilled into the walls of the
target hall, as described in \ref{TgtRmLt}, 
and shown in Fig.~\ref{target_room}. 

\begin{figure}[tb!]
\centering
\includegraphics[width=3.0in]{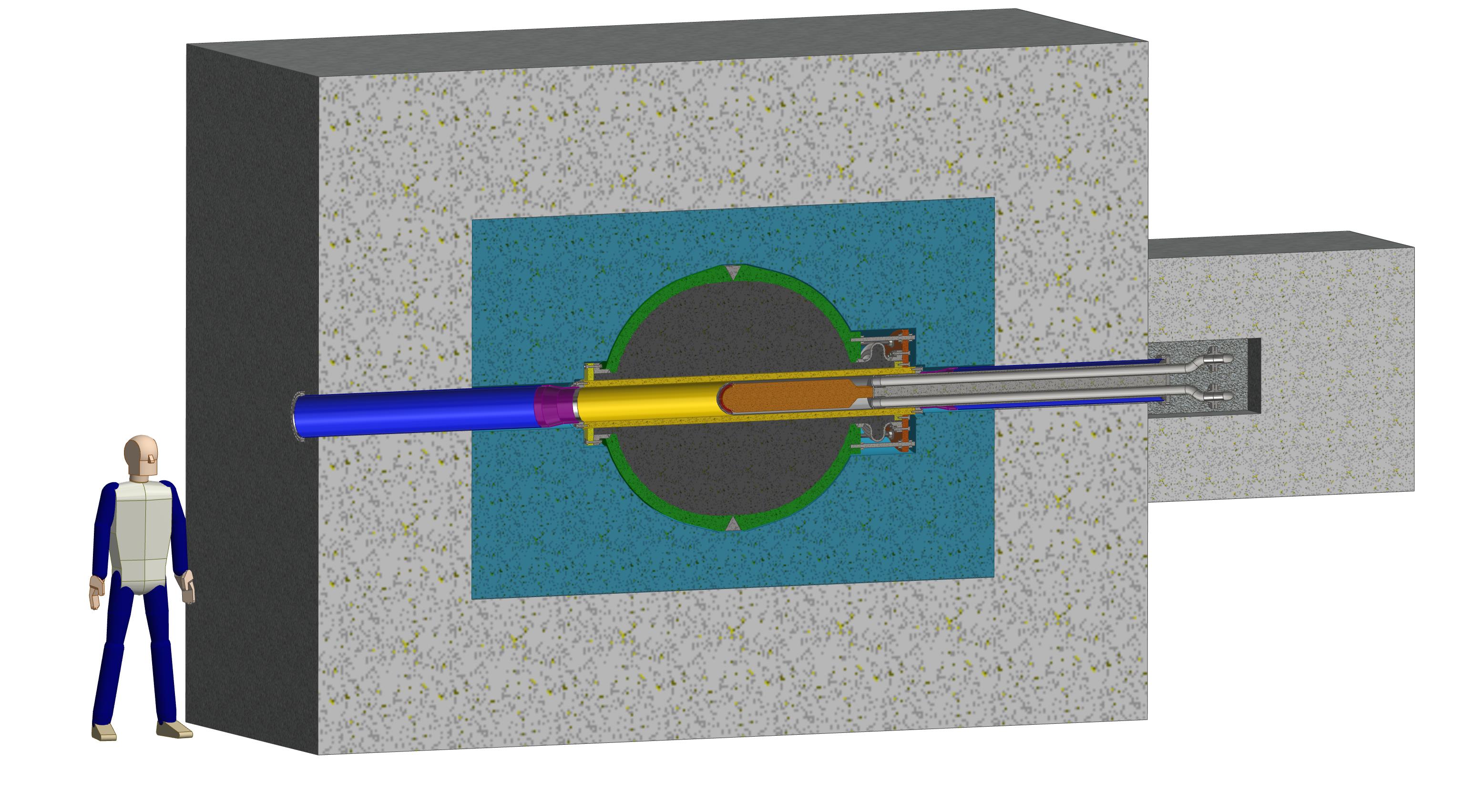}
\caption{{\footnotesize Section through the target system and the surrounding shielding. The target torpedo is surrounded by a Li/Be  sleeve pressure vessel. The shielding consists of inner layers of steel (shown in blue) and outer layers of boron rich concrete (shown in grey).}
\label{shielding}}
\vspace{0.2in}
\end{figure}

\begin{figure}[b!]
\centering
\includegraphics[width=0.55\columnwidth]{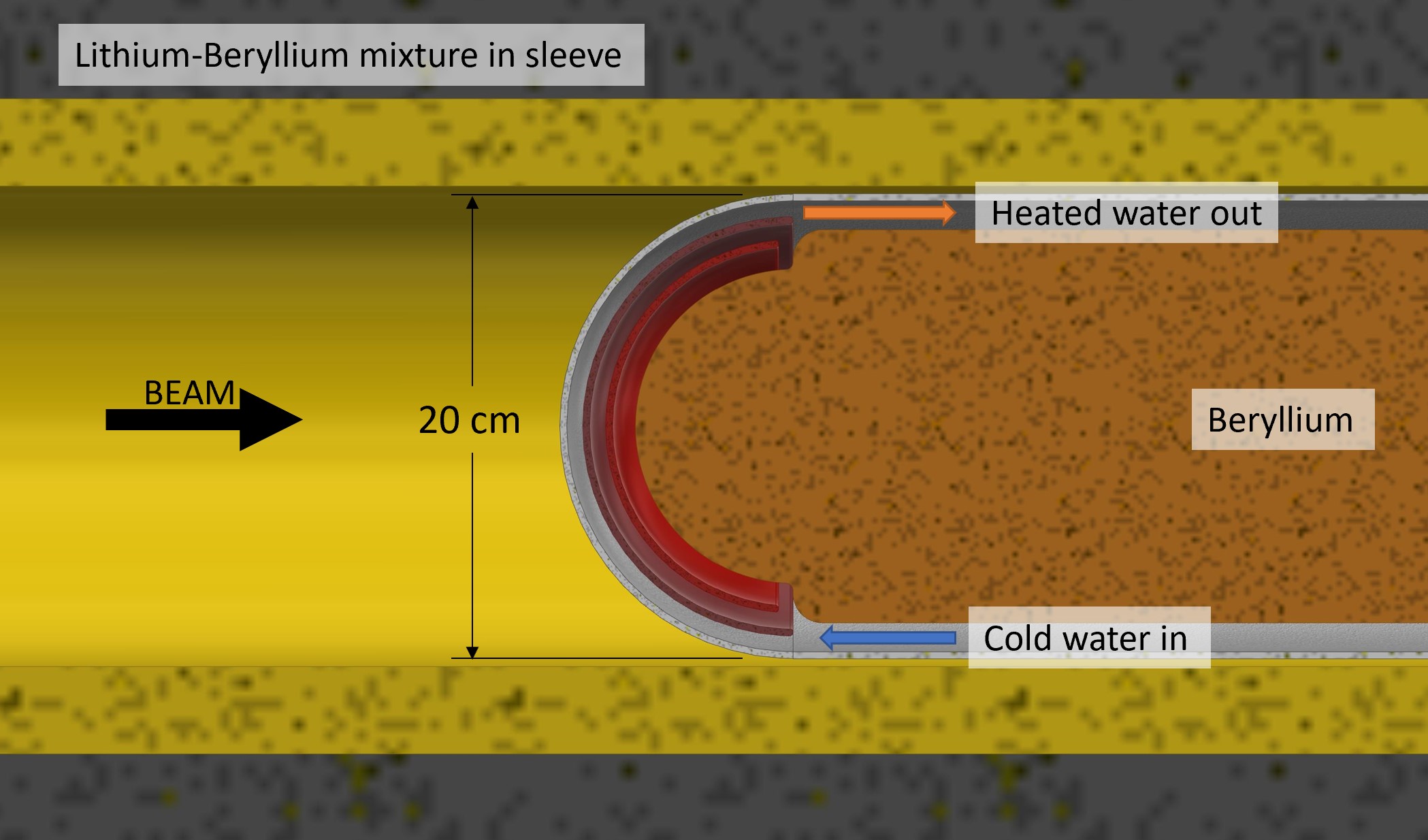}
\caption{{\footnotesize  Inside of the target vessel. The beam is spread out over the 
         face of the target by upstream wobbler magnets running at least at 50 Hz. 
         Cooling is provided by circulating heavy water that also serves to generate 
         more neutrons. Protons, with a range of about 2 cm, are stopped at various 
         depths  in this structure (depending on radial position). Neutrons stream
         into the sleeve surrounding the target, consisting of a mixture of highly 
         enriched ($>$99.99\%) $^7$Li and beryllium. $^8$Li is produced by neutron 
         capture.} \label{target-01}}
\vspace{0.2in}
\end{figure}

The head of the torpedo, the actual target struck by the protons,
consists of three nested beryllium hemispheres
each having a thickness of 3 mm with a gap of 7 mm of heavy water between them, 
shown in Fig.~\ref{target-01}. 
Heavy water is used as the cooling fluid, as the absorption cross section for neutrons is
 lower, and there are added neutrons from the breakup of the deuteron.  
% These increase the neutron flux reaching the $^7$Li in the sleeve.
GEANT4 calculations indicate an almost 40\% improvement in the $^8$Li yield for
D$_2$O over H$_2$O.
Using heavy water does introduce tritium issues, however, the completely-sealed design of the primary cooling loop assures that tritium will not be released into the environment.  Mitigation of tritium will be considered as part of a future decommissioning plan, where recovery and purification of the heavy water will be addressed. 

Water is circulated onto the target using a central divider plate made of 
beryllium, to separate the water flow.  
Water flows smoothly through the gaps between the beryllium hemispheres, 
Computational analyses of heat flow, temperature profiles and 
thermal stresses at the boundaries 
indicate acceptable performance of this target design, and show the target can handle the 600~kW of beam power.
\begin{figure}[t!]
\centering
\includegraphics[width=1.0\columnwidth]{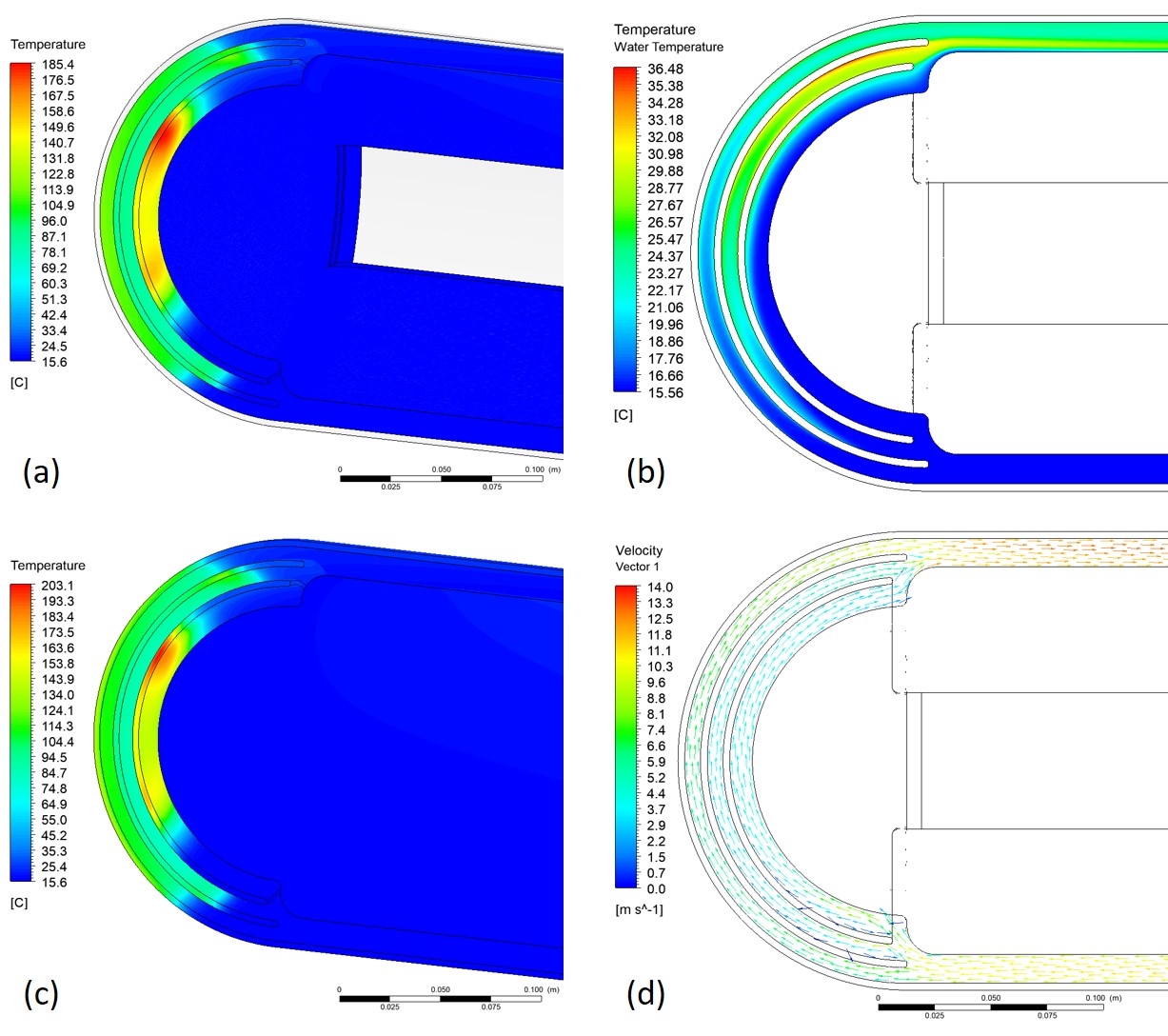}
\caption{{\footnotesize 
          (a) Temperature at the interface between beryllium and water.
          (b) Bulk water temperature away from the interface with the beryllium.
          (c) Beryllium temperature at the 500 GPM water flow rate.
          (d) Velocity vectors at the vertical mid-plane of the torpedo.}
\label{target_cfd_1}}
\end{figure}
The target geometry is still being optimized, specifically to address the known issue of
blistering of beryllium at the stopping point of the proton beam under 
intense bombardment due to the low solubility of hydrogen in beryllium~\cite{LENS}.
The beam was modeled with a uniform energy distribution across the face of the torpedo filling an 18 cm diameter circular area.  The target flows 500 gallons of water per minute, (0.031 m$^3$/sec), with an inlet temperature of 60\degree\,F (15.56\degree\,C). The pressure drop across the target at this flow rate is 30.4 psid (209.6 kPa). The Computational Fluid Dynamics (CFD) analysis was single phase only. Fig.~\ref{target_cfd_1}~(a) shows the temperature of the beryllium-water interface.

\begin{figure}[t!]
\centering
\includegraphics[width=1.0\columnwidth]{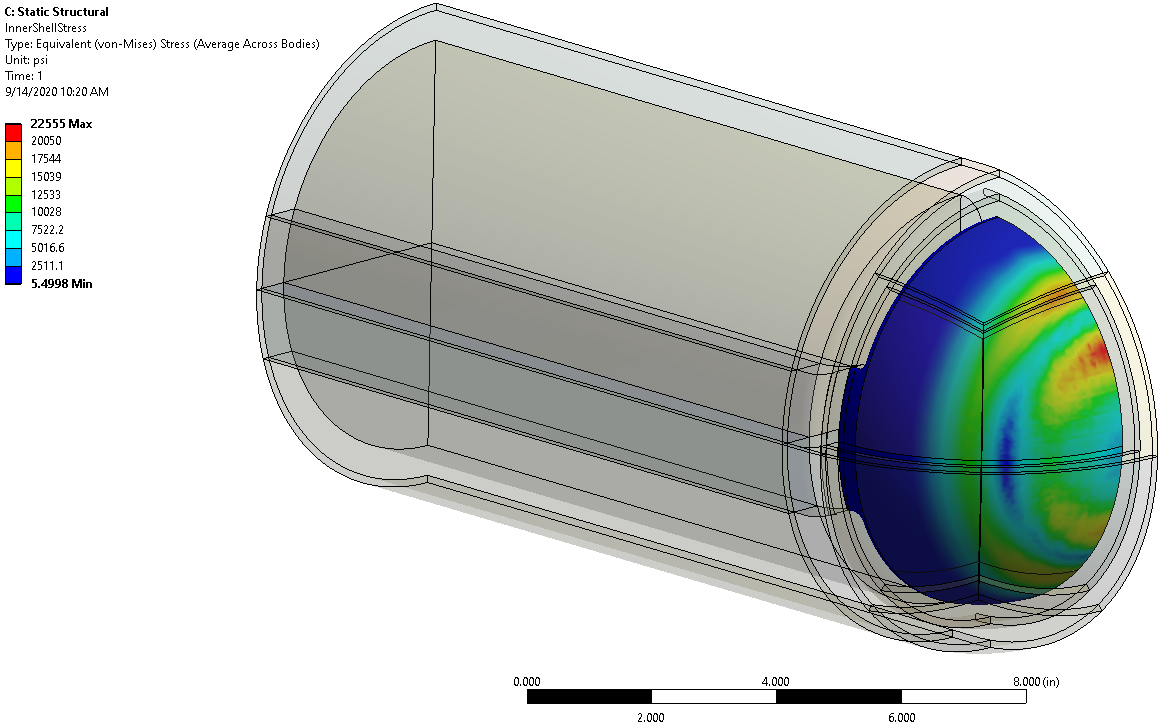}
\caption{{\footnotesize Thermal stress (Von Mises stress) on the innermost beryllium hemisphere.}
\label{target_cfd_2}}
\end{figure}

Fig.~\ref{target_cfd_1}~(b) shows the bulk water temperature away from the 
Be/water interface. Water temperatures are quite low away from the interface.
A future study will determine how far away the design is from the critical heat flux.
The interface temperature is higher than the boiling point of water at 1~atm, but 
the boiling temperature may be increased by increasing the pressure of the water. 
100~psi (689.4~kPa) static pressure was applied in this model. Two-phase analysis 
of the water cooling will happen once the actual energy distribution from the beam 
line is known.

Fig.~\ref{target_cfd_1}~(d) shows the velocity distribution in the water 
and shows good flow development between the beryllium hemispheres.
The single-phase analysis was primarily to look at the thermal stress in the 
beryllium as this was a problem in earlier designs. The single-phase temperature
distribution in the beryllium is more conservative because introducing boiling 
will reduce the beryllium temperature. Thermal stresses were analyzed per the 
methods of the ASME Boiler and Pressure Vessel Code Section 8, division 2.
Fig.~\ref{target_cfd_1}~(c) shows the temperature distribution in the beryllium.
Fig.~\ref{target_cfd_2} shows the maximum thermal stress in the beryllium,
which happens on the innermost hemisphere.  
The maximum stress there is 22.6~ksi (155.8~MPa) and the 
allowable stress is 36~ksi (248.2~MPa).  Stresses on the outer hemispheres are 
lower. This analysis showed that the target has sufficient cooling characteristics 
to allow the long-term survival of the beryllium to thermal stress.

The target torpedo is surrounded by the sleeve containing 25\% $^7$Li and 75\% $^9$Be. It is a roughly spherical pressure vessel
with an outer radius of 84 cm, and an inner radius of 74.4~cm.  
%Optimization studies have shown that a mixture of lithium and beryllium with a beryllium fraction mass of 75\%  considerably increases the neutron yield and therefore the neutrino production.
The sleeve is filled by pouring liquid lithium into the sleeve  pre-loaded
with beryllium powder.
The sleeve vessel is rated at 2500~psi, 
owing to the need to load the liquid lithium
under pressure because of the significant decrease
in volume between liquid and solid phases of the lithium. Not doing so 
would introduce unacceptable voids in the sleeve material.
\begin{figure}[t!]
\centering
\includegraphics[width=1.0\columnwidth]{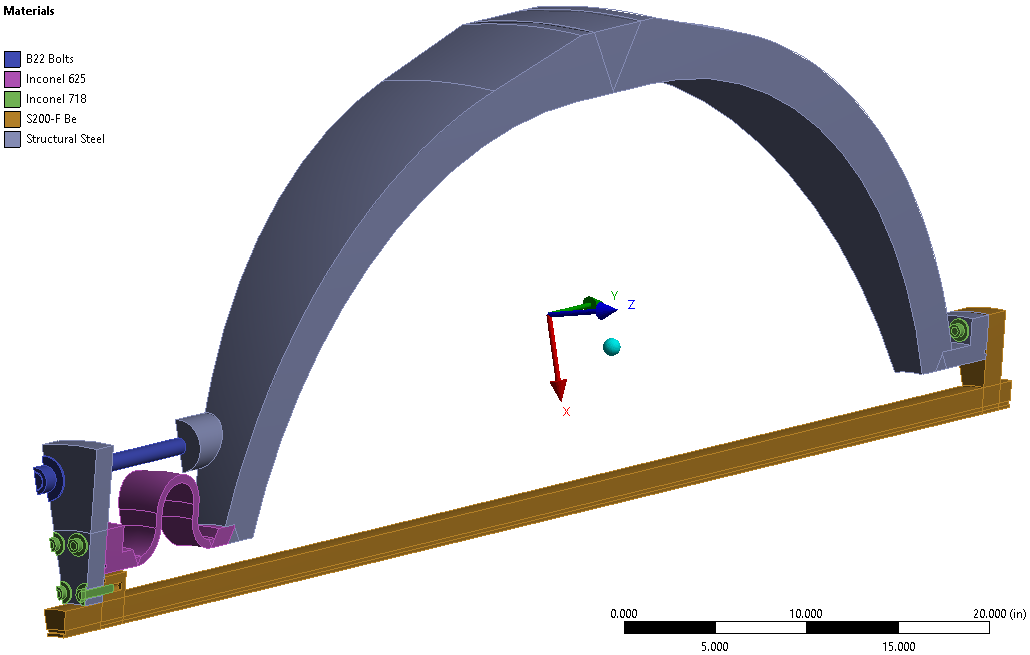}
\caption{{\footnotesize Model (15\degree symmetric wedge) of the sleeve as a pressure vessel showing the materials at different locations.}
\label{target_stress_1}}
\vspace{-5pt}
\end{figure}

The sleeve was also analyzed by the methods of the ASME Boiler and Pressure Vessel Code,
Section 8, Division 2. The materials in the model are shown in 
Fig.~\ref{target_stress_1}.  
The model showed that a plain carbon steel sleeve does not satisfy the ASME 
allowable stress. The sleeve material was changed to a structural steel, DIN K13049.  
Bolts are Inconel 718 and threaded rods in the model are B22 steel.
The analysis showed that the vessel satisfied the ASME stress allowable, but 
a peculiarity of the Code is that it is only concerned about whether a vessel 
is safe to internal pressure, not whether or not it leaks. It is imperative 
that molten lithium not leak out during the fill process, so much time was spent 
looking at the deformation of the o-ring grooves that seal the liquid lithium.

\begin{figure}[tb!]
\centering
\includegraphics[width=0.8\columnwidth]{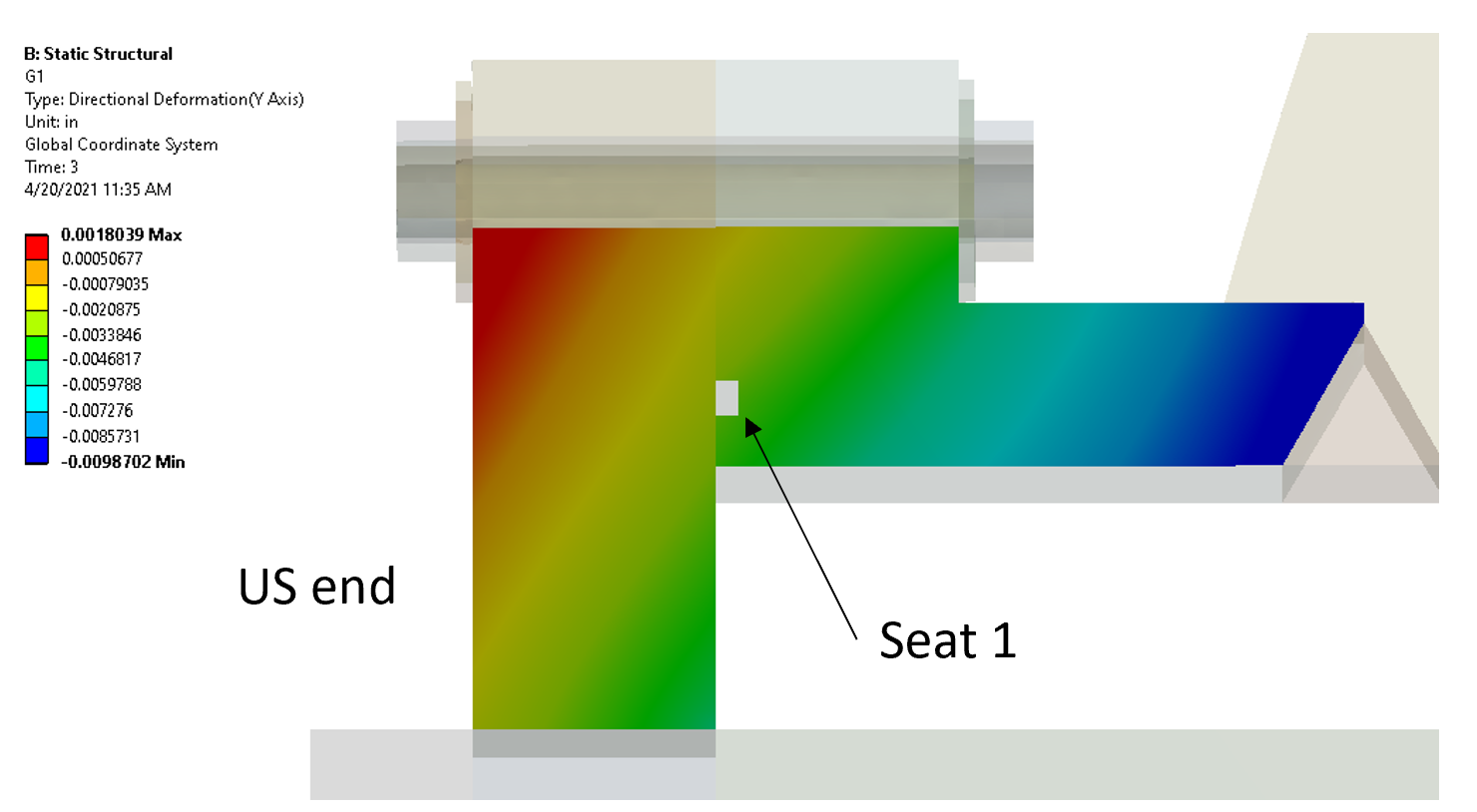}
\caption{{\footnotesize Deformation around the upstream (US) o-ring groove.}
\label{target_stress_2}}
\end{figure}
\begin{figure}[tb!]
\centering
\includegraphics[width=0.8\columnwidth]{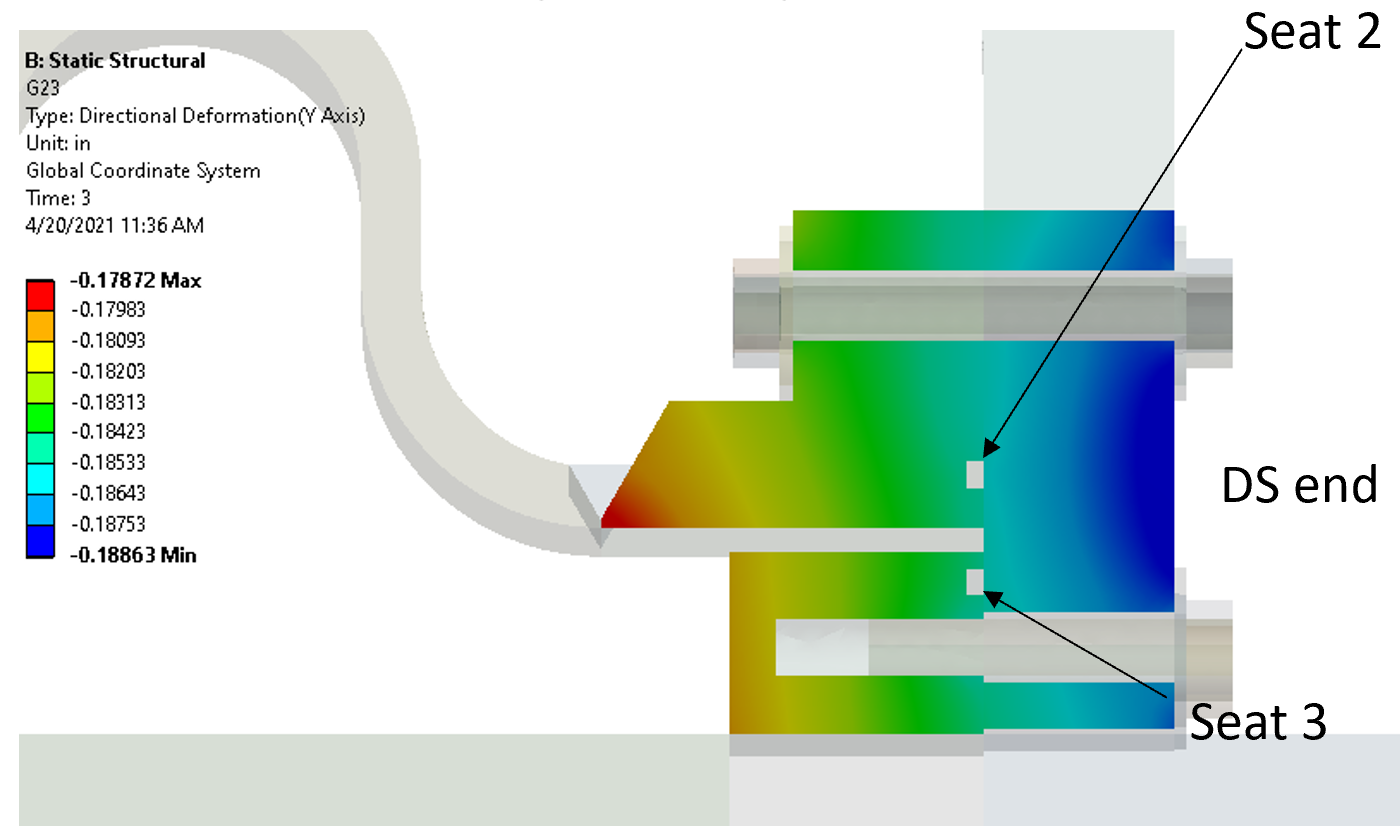}
\caption{{\footnotesize Seal configuration at the downstream (DS) end 
         showing deflections.}
\label{target_stress_3}}
\end{figure}
\begin{figure}[t!]
\centering
\includegraphics[width=1.0\columnwidth]{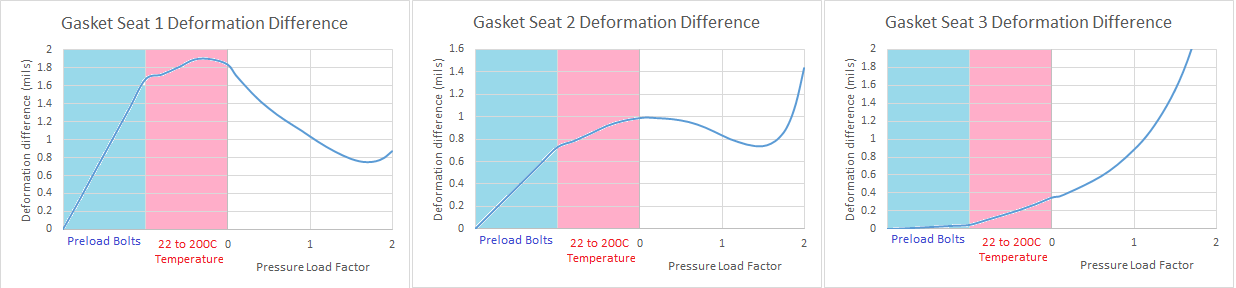}
\caption{{\footnotesize Gasket seat deformation differences (in thousandths of an inch,) as the loads are applied to the model.}
\label{target_stress_4}}
\end{figure}

The molten lithium is sealed in this design by custom Helicoflex metal seals.  
The seals crush by 0.021" (0.53~mm) minimum. Figs.~\ref{target_stress_2} and 
\ref{target_stress_3} show the seal
configuration. Fig.~\ref{target_stress_4} shows how the deformation of the
o-ring groove changes through the application of the load cases of the analysis.
Gasket seat 3 is of concern because as pressure is increased in the model, the 
deformation always increases.  Gasket seats 1 and 2 show a decrease in deformation 
as the pressure is applied.  In any case, the maximum deformation is about 0.001" 
(0.025~mm) at 2,500~psi (17.24~MPa), or 4.8\% of the seal crush.  
It is impossible to determine with certainty from these analyses whether the 
seals will leak, so prototyping the seals will happen in parallel with the 
development of the lithium injection machine.

% We conservatively assumed a $^{7}$Li isotopic purity of 99.99\% in our experimental 
% rate determinations. Although simulation studies 
% have shown that 99.995\% enrichment substantially improves the $^8$Li yield. 
% Even a very small amount of $^6$Li is damaging to $^8$Li production 
% because its thermal neutron capture cross section is a factor of 10$^3$ higher.
% We are told that the higher enrichment level is in fact achievable, and is the
% goal of several ongoing large-scale enrichment projects.
 
Beyond the sleeve is the shielding enclosure (cf. Fig.~\ref{shielding}), 
consisting of steel cylinders and plates, with a nominal thickness of 30 cm, 
and boron rich concrete of minimum thickness of 90~cm. Specifications and design 
details for this shield are addressed further in Section~\ref{shield-design}.

%\clearpage
\section{Installation at Yemilab}
\label{section:install}

Yemilab is an ideal location for the installation of IsoDAR for many reasons:   access to the underground site is possible with trucks to bring components directly from the surface;   ample space is available for an optimal layout for the experiment, allowing for separation of the cyclotron from the target by many meters -- such a layout allows for convenient installation and maintenance of equipment;  excavation of the cavern space fits in perfectly with the Yemilab construction timetable, in fact this construction is underway at present.
In addition,
the limestone rock environment is very dry and has a very low sodium content, reducing rock activation and eliminating need for ground-water management.   Finally, the LSC detector will have a sensitive liquid scintillator mass of 2.26 kilotons which allows for excellent physics reach.  
All these points will be (or have been in the case of the physics potential) elaborated further in this document.

We describe the cavern spaces in section~\ref{Yemilab_Description} and requirements for shielding in section~\ref{shield-design}.  Installation of IsoDAR is considered in section~\ref{section:underground} and requirements for electric power and other utilities in section~\ref{utilities}.

\subsection{Yemilab Description}
\label{Yemilab_Description}

\begin{figure}[t]
\centering
\includegraphics[width=0.6\textwidth]{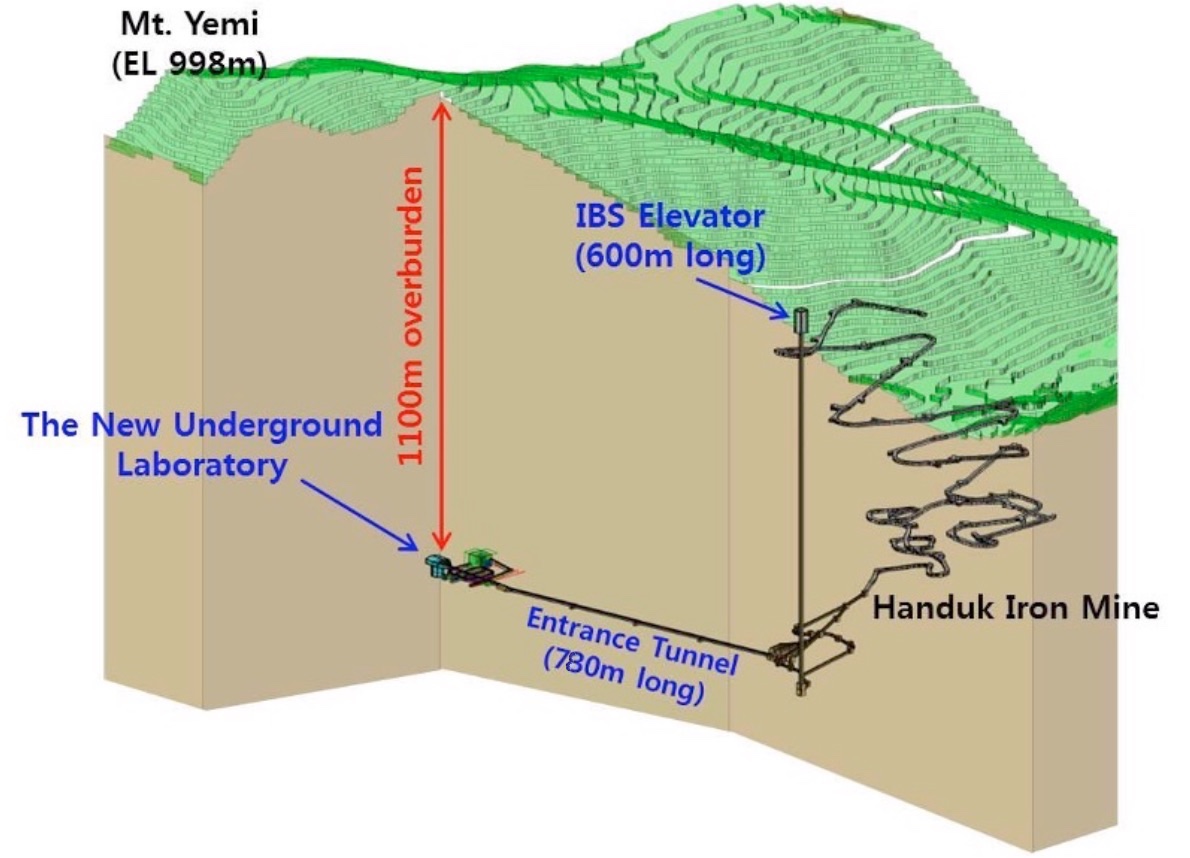}
\caption{{\footnotesize The location of Yemilab underneath Mount Yemi, adjacent to the Handuk Iron Ore Mine, about 200 km Southeast of Seoul.  Access to the Laboratory level is via the 6.6 km mine ramp.  Personnel access is also available using the elevator shaft.}
\label{YemiLayOut}}
\vspace{0.2in}
\end{figure}

\begin{figure}[t]
\centering
\includegraphics[width=4.in]{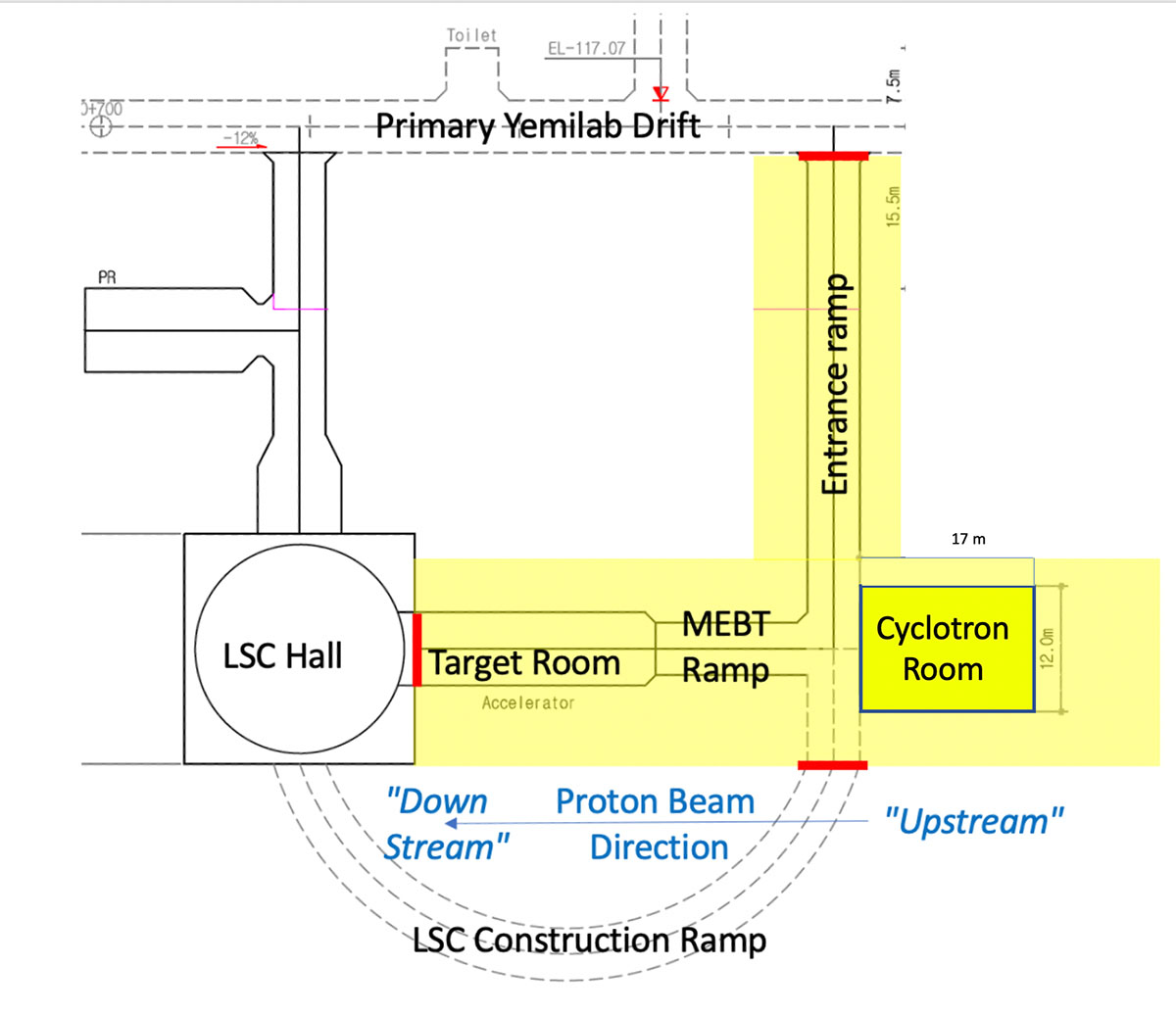}
\caption{\footnotesize  Layout of IsoDAR caverns at Yemilab.
These areas include the Entrance ramp, as the principal access to the IsoDAR experiment. This ramp drops about 6 meters
to the Cyclotron Room, then another 2 meters via the MEBT Ramp to the Target Room. The Target Room intercepts the LSC Hall at about the vertical half-way point.  All these ramps and rooms, except for the Cyclotron Room, are used for rock removal from the LSC Hall. 
\label{IsoYemiLayOut}}
\vspace{0.2in}
\end{figure}

The Yemilab complex is shown in 3-D in Fig.~\ref{YemiLayOut}. The Laboratory is located close to the Handuk Iron Mine, which has a truck-accessible ramp winding downwards for 6.6 km, and slope ranging between 12\% and 15\%.  From the base of this ramp (and also the terminus of a vertical personnel conveyance), the 780-meter long (12\% slope) Entrance Tunnel leads to the Laboratory.  
The Laboratory is a complex of about 9,000 m$^2$ with 26 laboratories and support rooms arranged in a ladder pattern. The Entrance Tunnel levels off into the Primary Yemilab Drift, the backbone for the system of crossing laboratory spaces. The cylindrical LSC Hall (20 meter diameter x 20 meter high) is close to the start of the backbone drift. This cylindrical hall is capped by a domed 22 x 22 meter cavern providing access to the top of the detector and housing electronics and infrastructure. The drifts built for excavation and rock removal from the LSC Hall will be the location for the IsoDAR equipment.

All Drifts and Ramps in the mine and Yemilab areas are nominally $5 \times 5$ meters, 
though some areas have more restricted cross sections 
because of ceiling-mounted ventilation trunks 
or other infrastructure needs. Designated Rooms are usually larger.

\subsubsection{Overview of IsoDAR Cavern Spaces}

\begin{figure}[t]
\centering
\includegraphics[width=4.in]{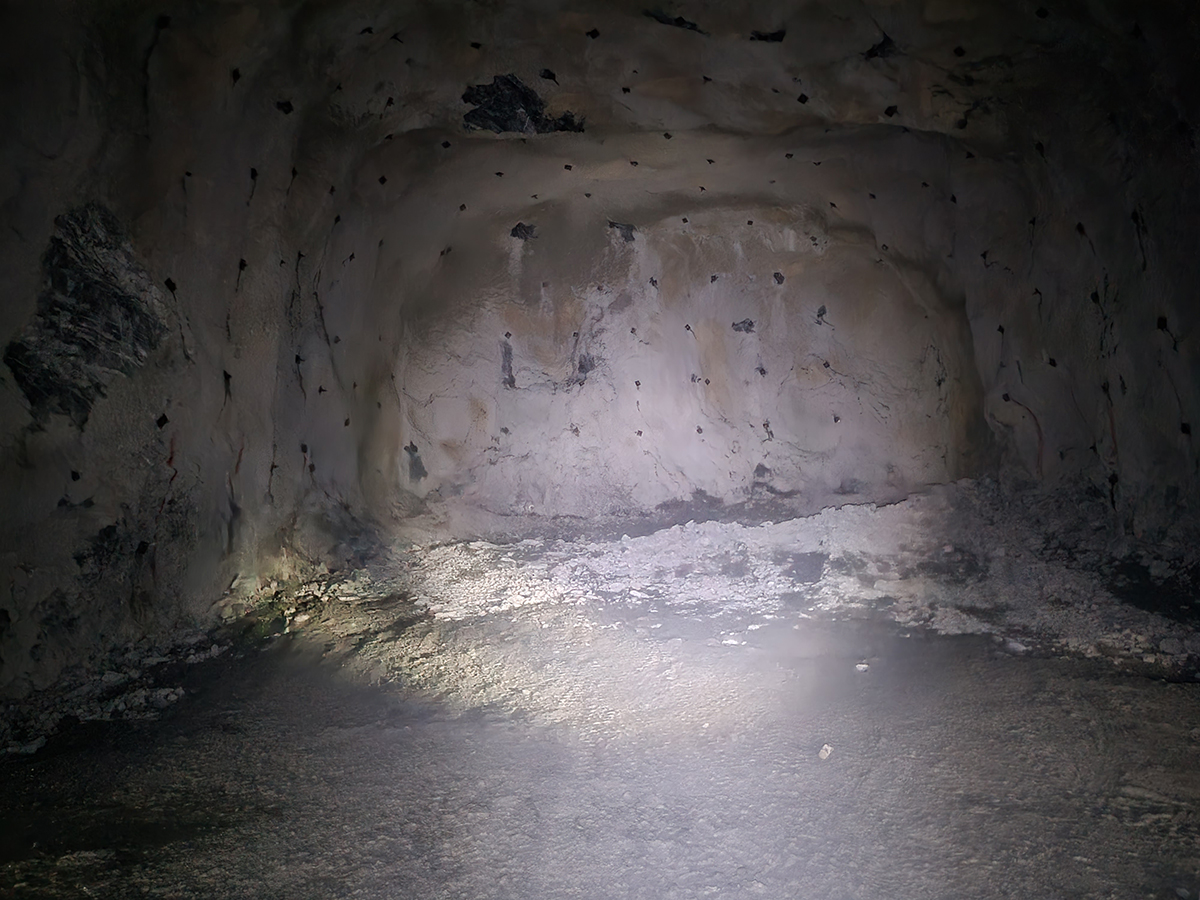}
\caption{\footnotesize  Completed Cyclotron Room excavation. 
\label{cycl-rm}}
\vspace{0.2in}
\end{figure}

The primary IsoDAR cavern spaces, shown in Fig.~\ref{IsoYemiLayOut}, 
consist of 
\begin{itemize}
    \item The Entrance Ramp -- 5 m (wide) $\times$ 5 m (high) $\times$ 45 m (long); 
    \item The Cyclotron Room -- 17 m (deep) $\times$ 12 m (wide) $\times$ 10 m (high); 
    \item The MEBT Ramp -- 5 m (wide) $\times$ 5 m (high) $\times$ 15.5 m (long); and     
    \item The Target Room -- 7 m (wide) $\times$ 7 m (high) $\times$ 22 m (long).
\end{itemize}
The cavern cross sections are all arched, or horse-shoe shaped, for rock stability: so the height in the Ramps and Target Room is the maximum height of a semicircular arch of radius 2.5 m (Ramps) and 3.5 m (Target Room).  The dome of the Cyclotron Room, as seen in Fig.~\ref{cycl-rm} is less arched, but has longer rock bolts.
The Ramps have 12\% slopes, while the floors of the Rooms are flat.  
The floor of the Cyclotron Room is about 6 meters below the level of the
Primary Yemilab Drift, the Target Room is another 2 meters lower.
The Target Room connects to the LSC Hall at the vertical mid-point of the LSC detector. The semicircular Construction Ramp allows removal of rock from the bottom level of the LSC Hall.  

All these spaces are required for the excavation of the LSC Hall, except for the Cyclotron Room, which is being excavated at the same time (Fig.~\ref{cycl-rm} shows the now-completed Cyclotron Room).  
The construction of the LSC Hall and all the IsoDAR caverns will be finished by the end of February, 2022.  
This is the last phase of major construction at the Yemilab site. Formal commissioning of the Laboratory and installation of the first experiments will follow.

It should be noted that large mining trucks are being used for removing
rock from the construction sites, and that these trucks can navigate all 
the turns and sharp corners in the ramp and tunnel system.  
These same trucks will be bringing cyclotron and other components into 
the IsoDAR area for assembly.

\subsubsection{The Cyclotron Room Layout}

\begin{figure}[t]
\centering
\includegraphics[width=4.in]{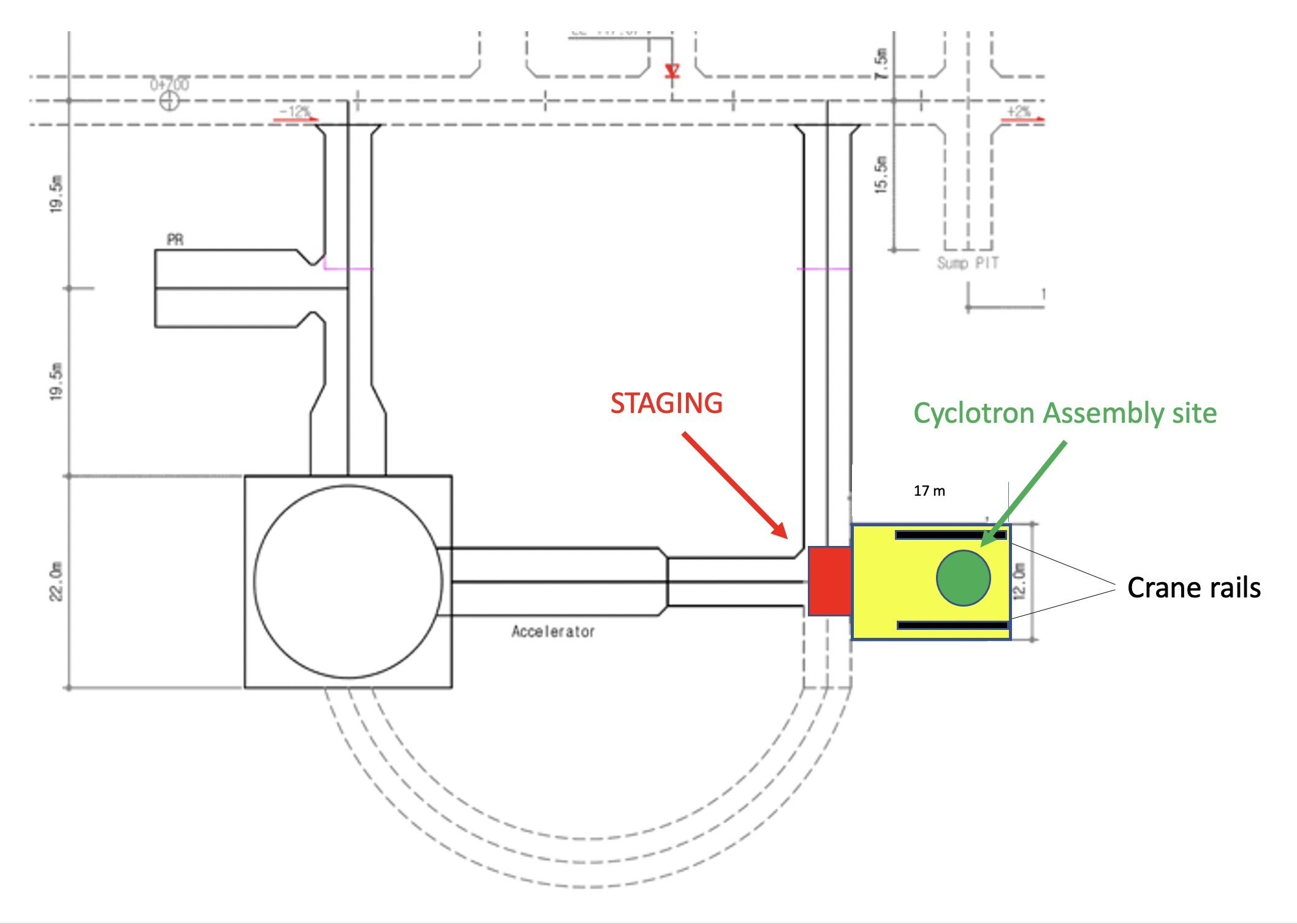}
\caption{{\footnotesize  Plan view of the Cyclotron Room.  Note the staging
area at the base of the Entrance Ramp, and the approximate location of the cyclotron. Also notice the rails for a bridge crane, highly valuable for the assembly of the cyclotron.}
\label{Cyclotron-staging}}
\vspace{0.2in}
\end{figure}

The size of the Cyclotron Room must be adequate both for operations needs 
and for assembly and installation of all the technical components.

The cyclotron is quite compact, at 6 meters' diameter.
There are some critical elements, such as the final RF amplifiers that need to be relatively
close by as the high-power lines between these and the cyclotron Dees must be as short
as possible.  Vacuum pumping equipment must also be nearby. 
Other elements can be located elsewhere, possibly even outside
the main IsoDAR areas for easy access and maintenance.  Cable trays will provide
the pathway for connection of these units and the cyclotron. 

But in the end, the size of the Cyclotron Room is set by the needs for assembly
of the cyclotron.  
Fig.~\ref{Cyclotron-staging} shows a conceptual configuration for the Cyclotron Room, including a staging area at the base of the Entrance Ramp where components can be taken off of the truck (using a fork lift, for example) and moved into a laydown area adjacent to the cyclotron assembly point.  
The cavern size allows for a minimum of 3 meter clearance all around the cyclotron, and also for a bridge crane to facilitate assembly.  The specification for this crane will be done in conjunction with the (as yet to be selected) cyclotron builder, and will be related to the maximum weight of component pieces of the cyclotron steel.

One additional feature of the Cyclotron Room is that 
the cavern floor must provide a foundation capable of supporting the 
450~ton weight of the 
cyclotron.  Thick ($\sim$10--20~cm), large ($\sim$1--2~m$^2$)
steel plates must be set in the floor under the cyclotron support points to distribute the load.

\subsubsection{The Target Room Layout}
\label{TgtRmLt}

\begin{figure}[t]
\centering
\includegraphics[width=4.in]{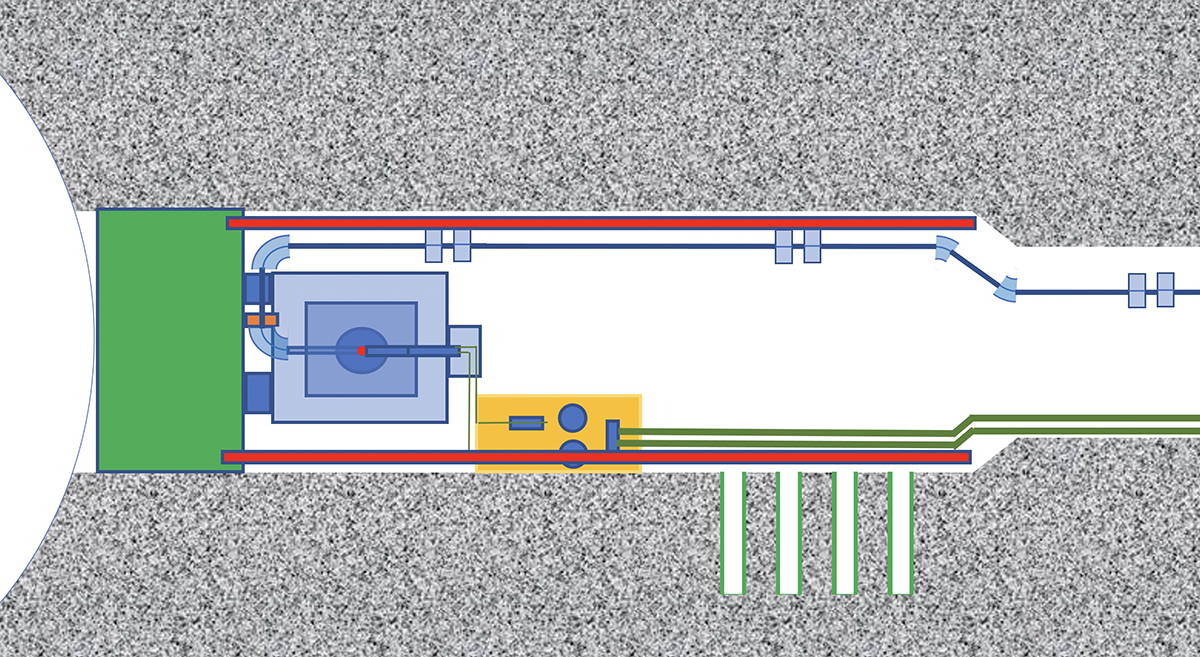}
\caption{{\footnotesize  Plan view for the conceptual layout of the Target Room. The MEBT beamline is seen along the top edge of the plan.  Beam enters from the right and is directed by two 90\degree bends to the target at the center of the large shielding block  at left, in blue, next to the shielding wall (indicated in green) designed to prevent neutrons from reaching the LSC.  The orange rectangle indicates a utilities-skid.  Red lines indicate crane rails. Bore-holes for spent targets are shown at lower right.    See text for further explanation. 
 }
\label{target_room}}
\vspace{0.2in}
\end{figure}

A conceptual layout of the Target Room is shown in Fig.~\ref{target_room}.  This
figure shows the beam line bringing the 60 MeV protons to the target in the center of the large shielding block
of iron and boron-loaded concrete.  

 The geometry pictured provides a solid angle of $\sim$0.2$\pi$ steradians for neutrinos produced in the sleeve,
or roughly a 5\% efficiency for the $\bar \nu_e$'s produced to reach the fiducial volume of the detector.
Maximizing the solid angle, i.e. the experimental efficiency, requires placing the target as close as possible to the detector.  However, adequate shielding must be provided to ensure neutrons do not reach the fiducial volume.  The geometry for the experiment is a careful balance between these two factors.

Cooling for the target is provided by the pumps, filters and heat exchangers shown on the yellow skid in Fig.~\ref{target_room}. As the 
(heavy) water
circulating here comes directly from the very high radiation environment of the target area, 
it must be treated as RAW (RadioActive Water) with full containment.  
If major maintenance of these components is required,
the entire skid can be placed in a shielded container and removed to the surface to a suitable hot cell area
where repairs or component replacement can be safely conducted.
Heat exchangers on the right-most side of the skid transfer heat to the secondary water system pipes, shown in green. These heat exchangers also provide environmental isolation of the two water systems. The secondary water system, and eventual conduction of the heat to the surface, will be discussed below in the chapter on utilities.

An important element of the Target Cavern is provision for target changes. 
Over the 5 years of IsoDAR operation, targets will need to be replaced.   We are drawing on
experience at high-power-beam installations (for example, Fermilab, SNS, JPARC, ISIS after) in developing target-changing procedures.  Space within the hall is adequate for this operation.   
Spent targets must remain within the target space for several years to cool before transport to the surface.
For storage, we are proposing that a series of $\sim$25 cm diameter bore holes be drilled in the wall of the cavern.  These are
shown schematically in Fig.~\ref{target_room} as the four perforations in the lower wall, right.  A detailed study will establish the desired number of such storage bore holes.
The bore holes will be made deep enough to accommodate a steel plug of $\sim 0.5$ meter to block
the opening.

A medium-capacity bridge-crane system will ease construction and can be used during initial assembly, and for target-changes.
 The capacity of this will be in the 10-15 ton range.  
 The crane rails are shown in red in Fig.~\ref{target_room}.

\subsection{Shielding Considerations}
\label{shield-design}

As the neutrino flux delivered to the detector is directly related to the number of neutrons produced in the target and sleeve, management of these neutrons must be given close attention.  This relates to shielding, as well as environmental isolation of the IsoDAR area from the main Yemilab complex.  The latter is greatly facilitated by the single point of connection between the IsoDAR caverns and the body of the laboratory. The top of the entrance ramp will be sealed by a properly-designed door that is closed except to allow entry of personnel, probably through an airlock that allows a slight negative pressure to be maintained on the IsoDAR side.\footnote{A single egress point requires the installation of a Refuge Chamber in the Target Room for underground safety.  The number of persons allowed beyond the barrier door is limited to the capacity of this Refuge Chamber.} An extensive labyrinth will also be placed behind this door to prevent neutron migration.

Radiation levels when the cyclotron is delivering beam will be high, no personnel will be allowed inside the IsoDAR space under these conditions.  When the beam is turned off, residual radiation levels will be related to the amount of activation in the material and walls of the caverns caused by neutrons not stopped in the (neutron) shielding, and to the (gamma) shielding between these activated materials and the rest of the caverns.

\begin{figure}[t]
\centering
\includegraphics[width=4.in]{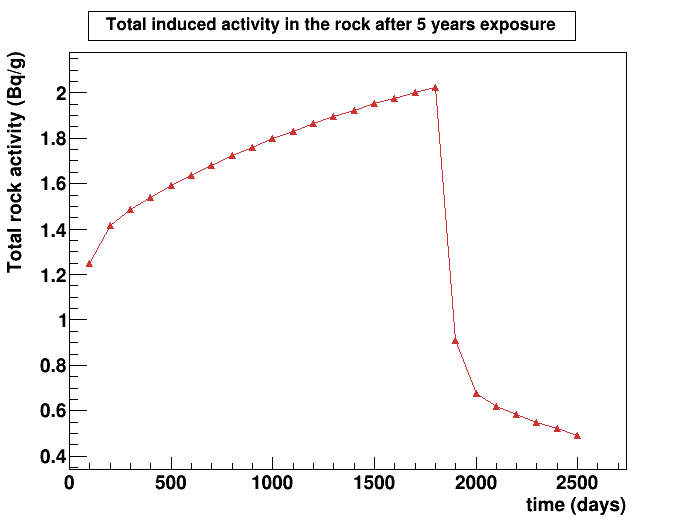}
\caption{{\footnotesize  The curve shows the buildup of activity in the surrounding rocks for the 5-year beam-on-target period.  Once beam is shut off after 1825 days, rapid decay of the many short-lived isotopes occurs, leaving only long-lived activities described in \ref{long-halflife}.  The "end of the experiment" is declared at approximately day 2500, by which time the activity level has dropped very significantly. The vertical scale should be viewed as arbitrary, it depends strongly on the composition of the rock.
 }
\label{rock-radioactivity}}
\vspace{0.2in}
\end{figure}

Shielding requirements can be divided into short-term (hours or days after beam shutoff), or long-term (years after last beam). The former relates to gamma radiation fields experienced by personnel entering for maintenance activities, the latter to the final decommissioning of the experiment, and ensuring that, after decommissioning,  radiation levels in the caverns are safely below Korean regulatory limits.  
As gamma radiation comes from decay of isotopes produced by neutron reactions, the critical consideration is the half-lives of these isotopes. Fig.~\ref{rock-radioactivity} traces a typical activation curve, showing the buildup of activity during the experiment for material exposed to neutrons, and after the beam is shut off, the decay of this activity level.  

\subsubsection{Long Term Activation Considerations} 
\label{long-halflife}
When evaluating residual radioactivity in the caverns, only the rock of the walls needs to be considered.  Components that are brought in that may be activated will be removed after the experiment for safe disposal elsewhere, only the rock will remain.

The tail in the curve of Fig.~\ref{rock-radioactivity} in the case of Yemilab is due to just a few isotopes with half-lives longer than a few years, primarily  $^{60}$Co (5.3 years), and $^{152,154}$Eu (13.5, 8.6 years), produced by thermal neutron capture; and $^{22}$Na (2.6 years) when high-energy neutrons are present (the (n,2n) reaction has an 11 MeV threshold).  Assays of Yemilab rock (basically pure limestone) from the vicinity of the LSC Hall have yielded very low concentrations of Co (< 10 ppm), and Eu (< 1 ppm), but far more important is that the sodium concentration is exceedingly low (0.02\%).

An extensive GEANT4 study has been performed (described in detail in Chapter 5 of the CDR), resulting in the design of the shielding structure around the target and sleeve (layers of steel and boron-loaded concrete), and an evaluation of residual rock activity at different depths from the rock face.  For the case described above, 2 years after the final shutoff of beam, the highest concentration of activity (seen directly above the center of the target, in the first 10 cm of rock depth) was less than 0.05 Bq/g.  The Korean regulatory requirement allows a maximum of 10 Bq/g.  The exceedingly low value is a direct result of the favorable concentration of sodium in the rock.  

\subsubsection{Short Term Activation Considerations}

When access is required for maintenance or repair, background radiation results from decay of isotopes with half-lives in the hour-range.  In accelerator environments, two isotopes stand out as major contributors:  $^{24}$Na (15.0 hour) and $^{64}$Cu (12.7 hour), both are produced by capture of slow neutrons, and emit gammas in the 0.5 to 5  MeV range.  Again sodium is the most damaging environmental component and, as indicated above, Yemilab native rock carries very low sodium content. It is important to ensure that material brought into the caverns, primarily for rock surface finishing (shotcrete) or as concrete shielding blocks, is also low in sodium.  Copper is present in all the cables and magnet windings, but these are localized sources, and if activated can be shielded with portable high-Z panels that can be brought in for this purpose. When access is required, a radiation technician will survey the caverns, identify areas requiring localized shielding, and establish when, and for how long, personnel may work in these areas.

\subsubsection{Neutron Background in the LSC}
\label{LSCshield}

Background events caused by fast neutrons (above 3 MeV) affect the sensitivity of the LSC.  Natural backgrounds, from spallation caused by high energy muons penetrating from the surface, natural U/Th/K concentrations in the environment of the detector, and impurities in the detector material itself, are well understood and characterizable.  It is important that the fast neutron flux penetrating from the Target Room that could reach the fiducial volume is kept well below these background levels. The thick steel and concrete wall, shown in Fig.~\ref{target_room} as the green structure blocking the area between the Target Room and the LSC Hall, must accomplish this.  The CDR document describes the state of design for this wall, it is at least 4 meters thick, primarily of steel but with a component of concrete to absorb slow neutrons that are not easily absorbed by the iron itself.

\subsubsection{Shielding in Other Areas}

Though the highest concentration of neutrons occurs at the target, beam loss at any location inside the cyclotron and in the transport line to the target will also produce neutrons. Though at much lower concentrations, these loss points must be understood and addressed for shielding design.

\begin{figure}[t!]
\centering
\includegraphics[width=0.45\textwidth]{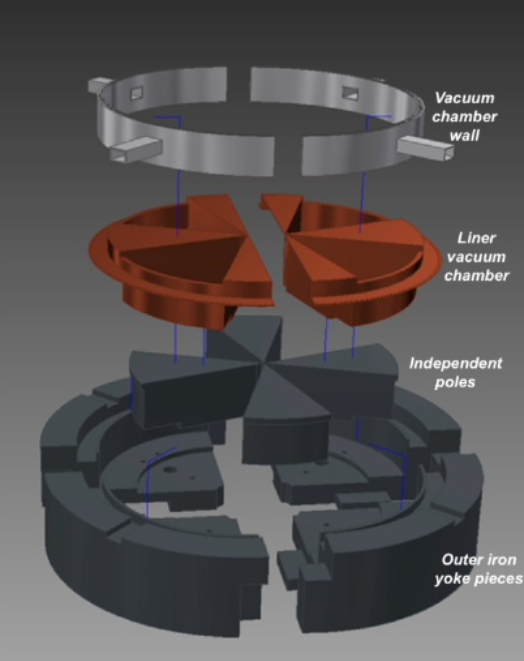}
\caption{{\footnotesize  Bottom half of cyclotron, split into parts.  The top half duplicates the liner (red) and magnet steel (grey). Not shown are the cylindrical coils, that fit between the inner ledge of the yoke and the outside of the four pole pieces. The steel yoke quarters are 45 tons each, so these may need to be assembled from smaller parts.}
\label{CyclotronParts}}
\vspace{0.2in}
\end{figure}

The cyclotron is basically a large block of steel, providing self-shielding for neutrons.  Maintenance inside the cyclotron that requires splitting the cyclotron apart will be supervised by a radiation technician who monitors the gamma levels and ensures personnel safety through proper localized shielding. Some areas of the cyclotron steel are thinner, and may require special localized shielding around the cyclotron to prevent fast neutrons from reaching the walls of the cavern. Further studies will be performed when the cyclotron design has been completed.

It is not possible to prevent all losses in the MEBT (Medium Energy Beam Transport) line, but they can be minimized, and the effects properly managed. These losses will come from scattering by residual gas in the vacuum pipes, and from growth of beam halo in the transport line.  The former is managed by transporting protons instead of the \htp ions extracted from the cyclotron (eliminating dissociation of the \htp ion), and specifying the highest practical vacuum in the beam line (<10$^{-5}$ Pa). The latter is managed by introduction of collimation stations that scrape the halo particles off of the beam.  These areas will receive appropriate shielding to contain neutrons produced.  Such areas are referred to as ``controlled'' loss points; inevitable beam loss is localized to areas that can be properly shielded. The point where the H$_2^+$ ions are stripped, close to the extraction point, is one such controlled loss point.  

When the ``source terms'' for these loss points are more completely understood, adequate shielding will be designed to satisfy the requirements for long and short term neutron control. These are standard procedures and we expect no difficulties.

\subsection{Installation Underground at Yemilab}
\label{section:underground}

The most challenging IsoDAR component for transport and installation will be the cyclotron.  Basically, it is a cylindrical block with a diameter of 6~m, 2~m height, weighing 450 tons, but Fig.~\ref{CyclotronParts} gives an idea of its constituent parts. These parts are transported separately and assembled on site.  The two particular challenges are the weight of the heaviest steel pieces of the cyclotron magnet, and the size of the magnet coils.

Commercial transport weight limits are around 50 tons per load, but the maximum weight for the magnet sections will be less than this, and will be set by the ability
to move heavy weights in the underground  caverns; in particular by the rigging capacity that will be available in the Cyclotron Room. The size of the steel for the magnet yoke and pole pieces can be adjusted to match this capacity, and a cost-benefit analysis must be performed.  Machining costs increase with number of pieces and tolerances of faces that must be machined with precision -- an optimal size for machining is about 45 tons. This is offset by the size and cost of crane or other rigging device that will be used for the precision assembly of the cyclotron pieces.  
Magnet steel pieces packed for shipping can be transported directly to the IsoDAR caverns underground.  This is facilitated by the presence of a rail terminal within a few km of the mine entrance, and by the capabilities of the large mining trucks to bring shipping pallets, transloaded from rail cars, down the mine ramp and directly to staging locations in the IsoDAR area.

\begin{figure}[t]
\centering
\includegraphics[width=3.in]{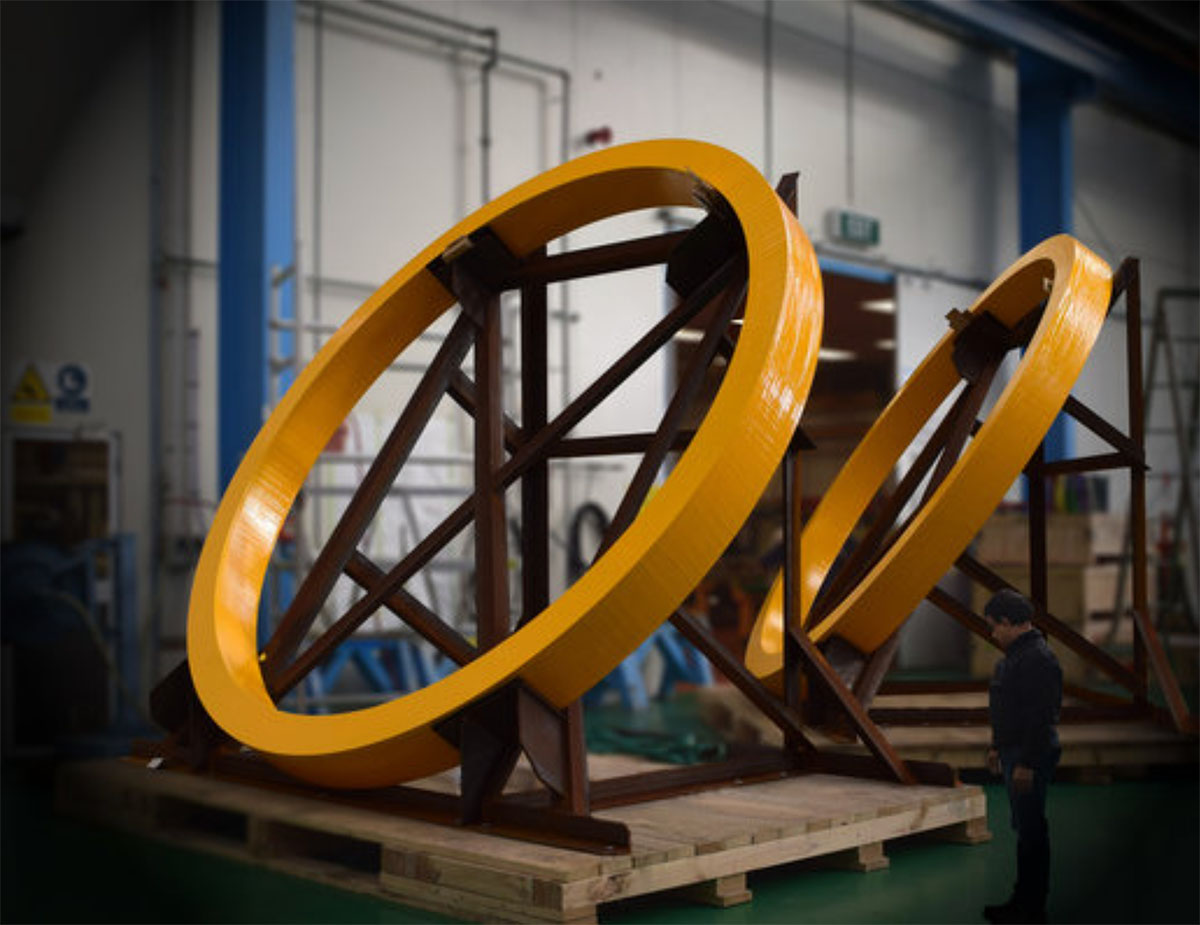}
\caption{{\footnotesize  Photo of cyclotron coils built by Buckley Systems, Auckland NZ.  These coils are 3.3~m diameter, ours are 4.95.  The proportions are about the same as ours. Notice the person in lower right.}
\label{coil}}
\vspace{0.2in}
\end{figure}

Although cyclotron coils only weigh about 1.2 tons each, they present a different challenge.  Each of the two copper coils is a torus of 4.95~m outer radius, and a cross section of 25 cm radially by 20 cm height. Fig.~\ref{coil} is included to give a sense of the coil size and shape. These cyclotron coils are 3.3~m in diameter, smaller than our 4.95~m.

The nominal dimension of the mine ramp and Yemilab drifts is 5 $\times$ 5~m, so in principle the coils, oriented at a 45\degree angle should fit through these passages.  However the mining trucks cannot be used for this, the shape of the bed does not allow for loading the coils in this fashion.  A special trailer, or a custom fixture for a fork lift, will need to be designed that can hold a coil oriented at the proper angle, along the diagonal of the tunnel. 

If it is not possible to transport the coils through the ramps, the coils will need 
to be either wound underground, or segmented into two halves. Both of these options 
are technically feasible but would increase cost.  
A segmented coil carries the requirement of splicing both ends of every turn to make 
a single continuous current path.  This drives the electrical properties from the most
efficient with several hundred turns, to a very high-current, low-voltage 
configuration required for a coil with no more than about 15 turns.

\subsubsection{Installation Sequence}

The first pieces to be installed will be the large iron and concrete shielding wall, 
shown as the large green block in Fig.~\ref{deployment} at the end
of the target hall.  
The total volume of
this wall is about 200 cubic meters, and has a weight of over 1,200 tons.
It will be important to ensure the rock at the lip of this cavern is properly reinforced
to support this weight.  It will need to be installed before the structure for the 
LSC is put in place, to allow rigging and installation of support structures from 
both sides. The design of this wall must ensure tight fits and minimize cracks, 
as neutrons are notorious for finding their way through even the narrowest spaces
between blocks.  

Next will be the cyclotron. Trucks will deliver components to the Staging Area shown 
in red at the bottom of the Entrance Ramp in Fig.~\ref{Cyclotron-staging}.
They will be offloaded by a fork lift, and moved to a location inside the Cyclotron 
Room where an overhead crane can perform the precision rigging needed for assembly 
of the cyclotron.

Following the cyclotron will be the target shielding and sleeve.  
As can be inferred from Fig.~\ref{shielding} this structure is made up of large 
blocks of special concrete and steel pieces that will be assembled using the overhead
crane.  The 20 cm diameter beam pipe running the length of this structure interfaces 
with the evacuated entrance beam line, and allows convenient insertion of the 
target torpedo from the back.

Then follow the beam line components, power supplies, and support equipment.  
The MEBT will be assembled last, as it crosses the entrance ramp
and so blocks truck access to the area.

\subsection{IsoDAR Utilities}
\label{utilities}

The utilities for IsoDAR (electricity, water and air) will be integrated with
substations and distribution centers designed to service the needs of the whole
laboratory. 

\subsubsection{Electric Power Delivery}

The power requirement for IsoDAR is 3.5~MW. 
The proton beam carries 600~kW of power. The efficiency of the accelerating 
system (wall-plug to beam through the RF system) is about 50\%, yielding
1.2~MW for the RF power alone.
The remaining electricity is for powering the ion source, cyclotron- 
and transport magnets, vacuum and water pumps, instrumentation, and controls.

KEPCO (Korean Electric Power Corporation) is currently contracted to deliver 
10~MW to the site. 7.5~MW is reserved for the use of the Handuk mine. 
Yemilab will require 2.5~MW in addition to the dedicated IsoDAR needs.  
To meet these IsoDAR needs the power system will have to be upgraded to a total of 
13.5~MW.
% This requires a new contract with KEPCO, possible upgrades of the 22.9 kV lines into the Handuk substation, and properly-sized cable running 3.3 kV power down the personnel shaft to the Yemilab substation.

\subsubsection{Cooling Water}

The 3.5 MW of electrical power supplied produces 3.5 MW of heat which needs to be removed.
This will  be accomplished with cooling water. No cooling water was originally 
planned for Yemilab, air circulation was sufficient to handle the expected Laboratory 
heat load.  Pumping cooled air into the main Laboratory may still be sufficient for 
the main body of Yemilab, but the IsoDAR heat load will require a surface-located,
modestly-sized cooling tower (<1000 ton), with water lines run up and down through 
the personnel shaft. 

Two ``Primary'' cooling loops are required for IsoDAR: the electrical systems having
``Low Conductivity'' water (LCW); while the target loop is filled with heavy water, 
this being part of the neutron-producing system as well as being the target coolant.   
Both these primary loops have heat exchangers that transfer their heat to a single 
secondary loop, which is run through the cooling tower on the surface.

\subsubsection{Air circulation}

Air into the main Yemilab area is provided by ducts running from the surface, down the personnel shaft and along the Yemilab Access Ramp.  It is released at selected end points in the Laboratory to ensure continuous airflow through the underground spaces.  As air is forced in, pressure in these Laboratory areas will always be slightly positive.

The IsoDAR areas should be maintained at a slightly negative pressure, so air would always be flowing into this area.  This is accomplished by drawing air out of the area and ducting it to the surface.  Air is replenished from the main Yemilab atmosphere, through the only portal entryway at the head of the IsoDAR entrance ramp.

\section{Conclusion}

This paper has reviewed the status of IsoDAR@Yemilab, demonstrating that feasible technical and infrastructure designs have been developed for all aspects of the experiment. 
This paper quotes and condenses the CDR~\cite{CDR},  which was developed within the context of reaching preliminary approval at Yemilab so that cavern construction could move forward.   
These designs show that IsoDAR@Yemilab can be constructed, and
perform beyond Standard Model searches that reach deeply into unexplored parameter space \cite{Isophysics}.

\section{Acknowledgments}
This work was supported by NSF grants PHY-1912764 and
PHY-1626069, and the Heising-Simons Foundation.
Winklehner was also supported by funding from the Bose Foundation.
We gratefully acknowledge support for this work from the Korean Institute for Basic Sciences, Grant IBS-R016-D1.

\clearpage
\bibliographystyle{unsrt}
%\bibliography{bibliography.bib}

\begin{thebibliography}{99}

\bibitem{LSC} Seon-Hee Seo, 
\newblock Neutrino Telescope at Yemilab, Korea.
\newblock  arXiv:1903.05368 [physics.ins-det]

\bibitem{PRL}    A.~Bungau, {\it et al.},
\newblock  Proposal for an Electron Antineutrino Disappearance Search Using High-Rate $^{8}$Li Production and Decay,
\newblock  Phys.\ Rev.\ Lett.\  {\bf 109}, 141802 (2012),
\newblock  arXiv:1205.4419 [hep-ex].
  %%

\bibitem{elastic}    J.~M.~Conrad, M.~H.~Shaevitz, I.~Shimizu, J.~Spitz, M.~Toups and L.~Winslow,
\newblock Precision $\bar{\nu}_e$-electron scattering measurements with IsoDAR to search for new physics,
\newblock  Phys.\ Rev.\ D {\bf 89}, no. 7, 072010 (2014),
\newblock  arXiv:1307.5081 [hep-ex].

\bibitem{CDR}  J.~R.~Alonso, J.~M.~Conrad, D.~Winklehner, S.~H.~Seo, K.~M.~Bang, Y.~D.~Kim, K.~S.~Park (for the IsoDAR Collaboration)
\newblock  IsoDAR@Yemilab: A Conceptual Design Report for the Deployment of the Isotope Decay-At-Rest Experiment in Korea's New Underground Laboratory, Yemilab.
\newblock (2021),
\newblock arXiv:2110.10635 [hep-ex].

\bibitem{Isophysics}	J.~R.~Alonso, C.~A.~Arg\"uelles, J.~M.~Conrad, Y.~D.~Kim, D.~Mishins, S.~H.~Seo, M.~Shaevitz, J.~Spitz, D.~Winklehner,
\newblock Neutrino Physics Opportunities with the IsoDAR Source at Yemilab,
\newblock  arXiv:2111.09480 [hep-ex].

\bibitem{miniboone_new}
A.A. Aguilar-Arevalo \textit{et al}. [MiniBooNE Collaboration],
Phys. Rev. D \textbf{103} 052002 (2021).

\bibitem{MB_antinu}
A.A. Aguilar-Arevalo \textit{et al}. [MiniBooNE Collaboration],
Phys. Rev. Lett. \textbf{110} 161801 (2013).

\bibitem{lsnd}
C. Athanassopoulos et al., Phys. Rev. Lett. \textbf{75} 2650 (1995); \textbf{77} 3082 (1996); \textbf{81} 1774 (1998);
Phys. Rev. C. \textbf{54} 2685 (1996); \textbf{58} 2489 (1998); A. Aguilar et al., Phys. Rev. D \textbf{64} 112007 (2001).

\bibitem{reactor} G. Mention, M. Fechner, T. Lasserre, T. A. Mueller,
D. Lhuillier, M. Cribier, and A. Letourneau, Phys. Rev.
D \textbf{83} 073006 (2011).

\bibitem{source}
C. Giunti and M. Laveder, Phys. Rev. C \textbf{83} 065504
(2011).

\bibitem{2109.11482} V.V.~Barinov, \textit{et al.} [BEST Collaboration], arXiv:2109.11482 [nucl-ex].

\bibitem{IceCube}
M.G. Aartsen \textit{et al.} [IceCube Collaboration],
Phys. Rev. Lett. \textbf{125} 141801 (2020).

\bibitem{global_reactor}
B.C Cañas, E.A. Garcés, O.G Miranda, M. Tortola,  and J.W.F. Valle,
Phys. Lett. B \textbf{761} 350 (2016). 

\bibitem{MarjonThesis}  Marjon Moulai, 
\newblock Light, Unstable Sterile Neutrinos: Phenomenology, a Search in the IceCube Experiment, and a Global Picture,
Marjon Moulai, 
\newblock arXiv:2110.02351 [hep-ex].
%0 citations counted in INSPIRE as of 11 Oct 2021

\bibitem{MHneutrons}  M. Hostert, D. McKeen, M. Pospelov, and N. Raj, 
\newblock Probing dark sectors with neutron-shining-through-a-wall experiments, 
\newblock in preparation.
  
\bibitem{costeffective}
A. Adelmann et al.,
\newblock Cost-effective Design Options for IsoDAR.
\newblock arXiv: 1210.4454 [physics.acc-ph], 2012.

\bibitem{EJNMMI}
L.H. Waites, J.R. Alonso, R. Barlow, J.M. Conrad,
\newblock What is the Potential Impact of the IsoDAR Cyclotron on Radioisotope Production:  a Review
\newblock EJNMMI Radiopharmacy and Chemistry 5, 6 (2020)

\bibitem{bungau:shielding}
Adriana Bungau, Jose Alonso, Larry Bartoszeck, Janet M. Conrad, Edward Dunton, Michael Shaevitz.
\newblock The Shielding Design Concept for the IsoDAR Neutrino Target.
\newblock Journal of Instrumentation {\bf 15}, T07002 (2020)

\bibitem{CDR2015}
M~Abs, A~Adelmann, JR~Alonso, S~Axani, WA~Barletta, R~Barlow, L~Bartoszek,
  A~Bungau, L~Calabretta, et~al.
\newblock {IsoDAR@ KamLAND: A Conceptual Design Report for the Technical
  Facility}.
\newblock {\em arXiv preprint arXiv:1511.05130}, 2015.

\bibitem{winklehner:nima}
Daniel Winklehner, Jungbae Bahng, Luciano Calabretta, Alessandra Calanna, Alok
  Chakrabarti, Janet M. Conrad, Grazia D'Agostino, Siddharta Dechoudhury,
  Vaishali Naik, Loyd Waites, et~al.
\newblock High intensity cyclotrons for neutrino physics.
\newblock {\em Nuclear Instruments and Methods in Physics Research Section A:
  Accelerators, Spectrometers, Detectors and Associated Equipment},
  907:231--243, 2018.

\bibitem{winklehner:mist1}
Daniel Winklehner, Janet Conrad, Joseph Smolsky, and Loyd Waites.
\newblock High intensity {H}{$_2$}{$^+$} beams from a
  filament-driven multicusp ion source.
\newblock {\em arXiv:2008.12292 [physics]}, August 2020.
\newblock arXiv: 2008.12292.

\bibitem{winklehner:RFQDIP}
Daniel Winklehner, Robert Hamm, Jose Alonso, Janet M. Conrad, Spencer Axani.
\newblock Preliminary Design of an RFQ Direct Injection Scheme for the IsoDAR High Intensity H{$_2$}{$^+$} Cyclotron.
\newblock Review of Scientific Instruments {\bf 87}, 02B929 (2016)

\bibitem{bungau:optimization}
Adriana Bungau, Jose Alonso, Larry Bartoszek, Janet M. Conrad, Michael Shaevitz.
\newblock Optimisation of the $^8$Li yield for the IsoDAR Neutrino Experiment.
\newblock Journal of Instrumentation {\bf 14}, P03001 (2019)
\newblock arXiv: 1805.00410.

\bibitem{winklehner:ICIS2021}
Daniel Winklehner, Janet M. Conrad, Joseph Smolksy, Loyd Hoyt Waites, Philip Weigel.
\newblock New Commissioning Results of the MIST-1 Ion Source.
\newblock Proceedings of the 2021 International Conference on Ion Sources, TRIUMF, September 2021.
\newblock https://indico.cern.ch/event/1027296/contributions/4479729/attachments/2312022/3934761/
ICIS21$\_$MIST-1$\_$Slides.pdf


\bibitem{winklehner:spiral}
Daniel Winklehner, Andreas Adelmann, Achim Gsell, Tulin Kaman, Daniela Campo.
\newblock Realistic Injection Simulations of a Cyclotron Spiral Inflector using OPAL.
\newblock arXiv: 1612.09018 [physics.acc-ph] (2016).

\bibitem{yang:vortex}
J.~Yang, A.~Adelmann, M.~Humbel, M.~Seidel, T.~Zhang, et~al., {Beam dynamics in
  high intensity cyclotrons including neighboring bunch effects: Model,
  implementation, and application}, Physical Review Special Topics-Accelerators
  and Beams 13~(6) (2010) 064201.

\bibitem{stetson:vortex}
J.~Stetson, S.~Adam, M.~Humbel, W.~Joho, T.~Stammbach, {THE COMMISSIONING OF
  PSI INJECTOR 2 FOR HIGH INTENSITY, HIGH QUALITY BEAMS}, in: {Proceedings of
  the 13th International Conference on Cyclotrons and their Applications},
  1992.

\bibitem{stammbach:vortex}
T.~Stammbach, S.~Adam, T.~Blumer, D.~George, A.~Mezger, P.~Schmelzbach,
  P.~Sigg, {The PSI 2mA beam and future applications}, in: AIP Conference
  Proceedings, Vol. 600, AIP, 2001, pp. 423--427.


\bibitem{winklehner:60mev}
D.~Winklehner, A.~Adelmann, J.M.~Conrad, S.~Mayani, S.~Muralikrishnan, D.~Schoen, 
M.~Yampolskaya
\newblock Order of Magnitude Beam Current Improvement in Compact Cyclotrons
\newblock arXiv:2103.09352 [physics] (2021)


  \bibitem{seidel:extraction}
  Injection and Extraction in Cyclotrons
  M. Seidel.
  \newblock Proceedings of the CAS--CERN Accelerator School:  Beam Injection, Extraction and Transfer, 
  Erice, Italy, 10-19 March 2017, edited by B Holzer, CERN Yellow Reports:  School Proceedings, Vol. 5/2018, CERN-2018-008-SP (CERN, Geneva, 2018).

\bibitem{LENS}
T.~Rinckel, D.V.~Baxter, J.~Doskow, P.E.~Sokol, T.~Todd.
\newblock Target Performance at the Low Energy Neutron Source.
\newblock Physics Procedia {\bf 26} (2012) 168-177.



%%\bibitem{Injector2}
%A.M.~Kolano, A.~Adelmann, R.~Barlow, C.~Baumgarten.
%\newblock A Precise Beam Dynamics Model of the PSI Injector 2 to Estimate the Intensity Limit
%\newblock Proceedings of IPAC2014, Dresden, Germany, TUPRI031, JACow Publishing (2014)

\end{thebibliography}
%\input{ConventionalFacilitiesCDRbib_v1.tex}

\end{document}